\documentclass[aps,prb,twocolumn,amsmath,amssymb,superscriptaddress,longbibliography]{revtex4-1}
\usepackage{graphicx}
\usepackage{dcolumn}
\usepackage{bm}
\usepackage{amsmath}
\usepackage{color}

\newcommand{\degCC}{\ensuremath{^{\circ}}}

\begin{document}

\title{
Recent achievements in \emph{ab initio} modelling of liquid water
}

\author{Rustam Z. Khaliullin}
\email{rustam@khaliullin.com}
\affiliation{
Institute of Physical Chemistry, Johannes Gutenberg University Mainz, Staudingerweg 7, D-55128 Mainz, Germany
}
\author{Thomas D. K\"uhne}
\email{kuehne@uni-mainz.de}
\affiliation{
Institute of Physical Chemistry, Johannes Gutenberg University Mainz, Staudingerweg 7, D-55128 Mainz, Germany
}
\affiliation{
Center for Computational Sciences, Johannes Gutenberg University Mainz, Staudingerweg 7, D-55128 Mainz, Germany
}
\date{\today}

\begin{abstract}
The application of newly developed first-principle modeling techniques to liquid water deepens our understanding of the microscopic origins of its unusual macroscopic properties and behaviour. Here, we review two novel \emph{ab initio} computational methods: second-generation Car-Parrinello molecular dynamics and decomposition analysis based on absolutely localized molecular orbitals. We show that these two methods in combination not only enable \emph{ab initio} molecular dynamics simulations on previously inaccessible time and length scales, but also provide unprecedented insights into the nature of hydrogen bonding between water molecules. We discuss recent applications of these methods to water clusters and bulk water.
\end{abstract}

\maketitle
\tableofcontents

\section{Introduction}

Liquid water is of paramount importance for life on Earth. That is why its properties and behaviour has been a subject of scientific investigation for several centuries~\cite{a:kauzmann}. It has long been established that liquid water has numerous unusual properties and exhibits anomalous behaviour for a wide range of different conditions. Nevertheless, despite extensive research, the underlying physical origins of many of these phenomena remain unclear. 


An isolated water molecule in the gas phase obeys a simple geometry and electronic structure. However, in the liquid phase, each water molecule forms multiple remarkably strong hydrogen bonds (HBs) with its neighbours and, thus, becomes a structural unit of the extended HB network. The complex structure, energetics and dynamics of this fluctuating network determine many anomalous macroscopic properties of liquid water. Therefore, a detailed investigation of elementary molecular processes in the HB network -- vibrations, reorientations, diffusion, HB-rearrangements -- is crucial for unravelling water mysteries, for better understanding of its role in nature and, consequently, for its better utilisation in artificial applications.

Experimental investigations of fundamental processes in the HB network at the molecular level has become possible only in the last few decades with the advent of sophisticated spectroscopic probes with femtosecond time resolution. However, even the most advanced spectroscopic techniques measure only the spectroscopic response of the complex system of interconnected water molecules, which is often difficult to interpret in terms of the real-time molecular structures and rearrangements. A recent highly controversial interpretation of the X-ray spectra of liquid water as evidence for its ``chains and rings'' structure~\cite{a:nilsson} is perhaps the most dramatic illustration of the intrinsic ambiguities of spectroscopic analysis.

The inconclusiveness of spectroscopic measurements has made computational modelling an indispensable tool for obtaining intelligible information from intricate experimental data. \emph{Ab initio} molecular dynamics (AIMD)~\cite{a:thecpmd,a:PayneRMP,b:aimd}, in which the interactions are obtained from accurate electronic structure calculations, has become widespread in computational studies of liquid water~\cite{a:cpmdwater,a:sprik2,a:marx-water1,a:marx-water2,a:autoion,a:marx-water3}. However, despite constant advances in high-performance computing, the great computational cost of AIMD still imposes severe constraints on the length and time scales attainable in AIMD simulations. Because of such restrictions the results of calculations often contain finite-size errors and/or statistical uncertainties due to insufficient sampling, the magnitude of which is difficult to estimate accurately. Here, we review a novel computational approach to AIMD that accelerates simulations with hundreds of water molecules such that trajectories as long as a nanosecond can be generated. We demonstrate that this new approach, which combines the advantages of both Car-Parrinello and Born-Oppenheimer molecular dynamics (MD), allows us to predict the properties of liquid water which are difficult to obtain with less efficient conventional AIMD methods.

The thermodynamics and kinetics of elementary processes in the HB network of liquid water are inextricably connected to the strength of interactions between water molecules. Therefore, in addition to computationally efficient algorithms, there is considerable interest in developing accurate electronic structure methods, which provide physical insight into the nature of interactions that determine the strength of hydrogen bonding. In this review, we discuss a recently developed energy decomposition technique, based on absolutely localized molecular orbitals (ALMOs), and its applications to water clusters that reveal an interesting and somewhat unexpected view of the electronic origins of hydrogen bonding. We also show how the combined application of the novel AIMD and energy decomposition methods help to address controversial questions related to the symmetry and order of the HB in liquid water.

\section{Computational methods}

\subsection{Second-generation Car-Parrinello molecular dynamics}

Until recently, \emph{ab initio} MD has mostly relied on two general approaches: Born-Oppenheimer MD (BOMD) and Car-Parrinello MD (CPMD), each with its advantages and shortcomings. In BOMD, the electronic wave function energy is iteratively minimized for each MD time step, making this approach computationally rather expensive. The CPMD~\cite{a:thecpmd, a:hutterCPrev} approach bypasses the expensive iterative minimization by introducing the fictitious mass for the electronic degrees of freedom and a suitably designed "on-the-fly" scheme for their propagation. The fictitious mass has to be small enough to ensure that the electrons follow the nuclei adiabatically, very close to their instantaneous electronic ground state~\cite{a:pastore}. As a consequence, the maximum permissible integration time step in CPMD is significantly smaller than that of BOMD, thus limiting the simulation timescales.

The recently developed second-generation CPMD method combines the best of both approaches by retaining the large integration time steps of BOMD and, at the same time, preserving the efficiency of CPMD~\cite{a:2ndcpmd}. The propagation scheme in second-generation CPMD relies on the appropriate usage of the information about the electrons from the previous MD steps and, in contrast to CPMD, does not require a fictitious mass parameter to maintain the accuracy of the Born-Oppenheimer dynamics. The superior efficiency of this new approach, which, depending on the system, varies between one to two orders of magnitude, has been demonstrated for a large variety of different systems~\cite{a:caravati1,a:caravati3,a:camellone,a:cucinotta,a:luduena,a:luduenarep}.



\textbf{Propagation of the electronic degrees of freedom.} Within mean-field electronic structure methods, such as Hartree-Fock and Kohn-Sham density functional theory (DFT), the electronic wave function is described by a set of occupied molecular orbitals (MOs) $|\psi_{i}\rangle$ or by an associated idempotent one-electron density operator $\rho = \sum_i | \psi_{i} \rangle \langle \psi_{i} |$. In the second-generation CPMD method, the propagation of the electronic degrees of freedom is achieved by adapting the predictor-corrector integrator of Kolafa~\cite{a:kolafa1, a:kolafa2} to the electronic structure problem. Firstly, the predicted MOs at time $t_n$ are constructed in terms of the electronic degrees of freedom from the $K$ previous MD steps:  
\begin{eqnarray}
|\psi_{i}^{p}(t_{n}) \rangle &=& \sum_{m=1}^{K} \rho(t_{n-m}) |\psi_{i}(t_{n-1}) \rangle \alpha_{m} \label{ASPCpredictor} \\
\text{where}~\alpha_{m} &=& {(-1)^{m+1}} m \frac{\left( 2K \atop K-m \right)}{\left( 2K-2 \atop K-1\right)} \nonumber
\end{eqnarray}
Secondly, the orbitals are corrected by performing a single step $| \delta \psi_{i}^{p}(t_{n}) \rangle$ along the preconditioned electronic gradient direction computed with the orbital transformation method~\cite{a:ot}. This leads to the final updated orbitals:
%
\begin{eqnarray}
  |\psi_{i}(t_{n}) \rangle &=& (1 - \omega) |\psi_{i}^{p}(t_{n}) \rangle + \omega | \delta \psi_{i}^{p}(t_{n}) \rangle, \label{ASPCcorrector} \\
  \text{where}~\omega &=& \frac{K}{2K-1}~\text{and}~K \ge 2 \nonumber
\end{eqnarray}

The second-generation CPMD method avoids the self-consistency cycle entirely and obviates the need for the computational expensive diagonalization, replacing it with a fast orbital transformation step. The electron propagation scheme is rather accurate and time reversible up to $\mathcal{O}(\Delta t^{2K-2})$, where $\Delta t$ is the integration time step. It keeps the electronic degrees of freedom very close to the Born-Oppenheimer values and allows for $\Delta t$ to be as large as in BOMD. 

\textbf{Energy functional and nuclear forces.} The total energy $E^{PC}[\rho^{p}]$ is evaluated from both predicted and corrected orbitals and represents an approximation to the Harris-Foulkes functional~\cite{a:harris, a:foulkes}:
\begin{eqnarray}
  E^{PC}[\rho^{p}] &=& \sum_{i} \langle \psi_{i} | \hat{H}[\rho^{p}] |\psi_{i}\rangle - \frac{1}{2} \int{d\bm{r} \int{d\bm{r}' \frac{\rho^{p}(\bm{r}) \rho^{p}(\bm{r}')}{|\bm{r} - \bm{r}'|}} } \nonumber \\
  &-& \int{d\bm{r} \, v_{XC}[\rho^{p}] \rho^{p}} + E_{XC}[\rho^{p}] + E_{II}, \label{HarrisEnergy}
\end{eqnarray}
where $\rho^{p}(\bm{r})$ is the predicted electron density associated with $|\psi_{i}^{p}(t_{n})\rangle$, $\hat{H}[\rho^{p}]$ is the effective Hamiltonian operator, $v_{XC}[\rho^{p}]$ the exchange-correlation (XC) potential, whereas $E_{XC}[\rho^{p}]$ and $E_{II}$ are the XC energy and the nuclear Coulomb interaction, respectively. 


The nuclear forces are computed by evaluating the analytic energy gradient $\bm{F}_{I}^{PC} = \nabla_{\bm{R}_{I}}E^{PC}[\rho^{p}]$. However, since $\Delta \rho \equiv \rho - \rho^{p} \neq 0$, the usual Hellmann-Feynman~\cite{a:rpfeynman} and Pulay forces~\cite{a:PulayForces} have to be augmented by the following extra term:
\begin{eqnarray}
  - \int{d\bm{r} \, \left\{ \left[ \frac{\partial v_{XC}[\rho^{p}]}{\partial \rho^{p}} \Delta \rho + v_{H}[\rho^{p}] \right] \left( \nabla_{\bm{R}_{I}} \rho^{p} \right) \right\}}, \label{HarrisForce}
\end{eqnarray}
where $v_{H}[\rho^{p}]$ is the Hartree potential, while $\rho$ is the corrected density that corresponds to $| \psi_{i}(t_{n}) \rangle$.

\textbf{Modified Langevin equation.} Despite the close proximity of the electronic degrees to the Born-Oppenheimer surface, the employed propagation scheme introduces a small dissipative energy drift during long MD runs. 
However, it is possible to correct for that by devising a modified Langevin-type equation that in its general form reads as: 
\begin{eqnarray}
  M_{I} \Ddot{\bm{R}}_{I} = \bm{F}_{I}^{BO} - \gamma M_{I}\dot{\bm{R}}_{I} + \bm{\Xi}_{I}, \label{LangevinEq}
\end{eqnarray}
where  $M_{I}$ are the nuclear masses, $\bm{R}_{I}$ the nuclear coordinates (the dot denotes time derivative), $\bm{F}_{I}^{BO}$ the exact Born-Oppenheimer forces, while $\gamma$ a damping coefficient, and $\bm{\Xi}_{I}$ is an additive white noise, which must obey the fluctuation-dissipation theorem $\left< \bm{\Xi}_{I}(0) \bm{\Xi}_{I}(t) \right> = 2 \gamma k_{b} T M_{I} \delta(t)$ in order to sample the canonical distribution.

Presuming that the energy is exponentially decaying, which had been shown to be an excellent assumption~\cite{a:2ndcpmd,a:dai,a:hutterCPrev}, it is possible to rigorously correct for the dissipation, by modelling the nuclear forces arising from our dynamics as: 
\begin{eqnarray}
  \bm{F}_{I}^{PC} = \bm{F}_{I}^{BO} - \gamma_{D}M_{I}\dot{\bm{R}}_{I}, \label{TdKmodel}
\end{eqnarray}
where $\gamma_{D}$ is an intrinsic, yet unknown damping coefficient to mimic the dissipation. 

By substituting Eq.~(\ref{TdKmodel}) into Eq.~(\ref{LangevinEq}) the following modified Langevin-like equation is recovered: 
\begin{eqnarray}
  M_{I} \Ddot{\bm{R}}_{I} = \bm{F}_{I}^{PC} + \bm{\Xi}_{I} \label{ModLangevinEq}
\end{eqnarray}
In other words, if we would know the unknown value of $\gamma_{D}$, in spite of the dissipation, it is nevertheless possible to guarantee an exact canonical sampling of the Boltzmann distribution, by simply augmenting Eq.~\ref{TdKmodel} with $\bm{\Xi}_{I}$ according to the fluctuation-dissipation theorem. 
Fortunately, the intrinsic value of $\gamma_{D}$ does not need to be known \textit{a priori}, but inspired by ideas of Krajewski and Parrinello~\cite{a:krajewski}, can be determined in a preliminary run using a Berendsen-like algorithm~\cite{a:berendsen} in such a way that eventually the equipartition theorem $\left< \frac{1}{2} M_{I} \Dot{\bm{R}}_{I}^{2} \right> = \frac{3}{2} k_{B} T$ holds. 
Although this can be somewhat laborious, but once $\gamma_{D}$ is determined, very long and accurate simulations can be routinely performed at a greatly reduced computational cost. 

\subsection{Decomposition analysis based on absolutely localized molecular orbitals}

\textbf{Energy decomposition analysis.} One of the most powerful techniques that modern first-principle electronic structure methods provide to study and analyse the intermolecular bonding is the decomposition of the total molecular binding energy into physically meaningful components. Such methods have shown that the HB is a result of the delicate interplay between weak dispersive forces, electrostatic effects (e.g. charge-charge, charge-dipole and charge-induced dipole interactions) and donor-acceptor type orbital (i.e. covalent) interactions, such as forward- and back-donation of electron density between water molecules. The extent of the different modes of interactions determines the strength of the HBs in water clusters and condensed phases and, consequently, all their physical properties.

The need for physically reasonable and quantitative useful values of the energy components has resulted in numerous decomposition schemes which have been proposed since the early years of theoretical chemistry~\cite{a:saptrev,a:saptmolphys,a:km,a:rvs,a:rvsX,a:csov,a:csov1,a:neda,a:nedaX,a:nedaDFT,a:cafi,a:scccms,a:dnc,a:blweda,a:theeda,a:cta}. In this review, we discuss a novel energy decomposition analysis scheme based on absolutely localized molecular orbitals (ALMO EDA)~\cite{a:theeda,a:cta}.  Unlike conventional MOs, which are generally delocalized over all molecules in the system, ALMOs are expanded in terms of the atomic orbitals of only a given molecule~\cite{a:stoll,a:gia0,a:nagata0,a:khal}. Although ALMOs were originally used to speed up the calculation of SCF energies for large ensembles of molecules~\cite{a:khal}, they are now widely used in energy decomposition analysis (EDA)~\cite{a:blweda,a:theeda,a:cta}. It should be mentioned that since the introduction of ALMO EDA~\cite{a:theeda}, ALMO-based decomposition methods have been extended to many-determinant wave functions~\cite{a:almo-cc}. Nevertheless, the application of the most recent approach to water is still very limited and here we focus on the decomposition procedure based on the mean-field methods such as Hartree-Fock and density functional theory.

ALMO EDA separates the total interaction energy of molecules ($\Delta E_{TOT}$) into a frozen density component (FRZ), as well as polarization (POL) and electron delocalization (DEL) terms
%
\begin{equation}
	\Delta E_{TOT} = \Delta E_{FRZ} + \Delta E_{POL} + \Delta E_{DEL}.
\end{equation}

The frozen density term is defined as the energy change that corresponds to bringing infinitely separated molecules into the complex geometry without any relaxation of the MOs on the monomers, apart from energetic modifications associated with satisfying the Pauli exclusion principle: 
\begin{equation}
\Delta E_{FRZ} \equiv E(R_{FRZ}) - \sum_{x} E(R_x), 
\end{equation}
where $E(R_x)$ is the energy of the isolated molecule $x$ with its nuclei fixed at the geometry that it has in the system. $R_{FRZ}$ is the density matrix of the system constructed from the unrelaxed MOs of the isolated molecules.

The polarization energy is defined as the energy lowering due to the intramolecular relaxation of each molecule's ALMOs in the field of all other molecules in the system. The intramolecular relaxation is constrained to include only variations that keep MOs localized on their molecules, i.e.
\begin{equation}
\Delta E_{POL} \equiv E(R_{POL}) - E(R_{FRZ}), 
\end{equation}
where $R_{POL}$ is the density matrix constructed from the fully optimized (polarized) ALMOs. That is, the $R_{POL}$ state is the lowest energy state that can be constructed from completely localized MOs. ALMOs are not orthogonal from one molecule to the next and, therefore, the minimization of the electronic energy as a function of the ALMO coefficients differs from conventional SCF methods. Mathematical and algorithmic details of the self-consistent field procedure for finding the polarized state $R_{POL}$ have been described by many authors~\cite{a:stoll,a:gia0,a:nagata0,a:khal}.

The remaining portion of the total interaction energy, the electron delocalization (DEL) or charge-transfer (CT) energy term, is calculated as the energy difference between the state formed from the polarized ALMOs ($R_{POL}$) and the state constructed from the fully optimized delocalized MOs ($R_{SCF}$). Thus, 
\begin{equation}
\Delta E_{DEL} \equiv E(R_{SCF}) - E(R_{POL}), 
\end{equation}
where $\Delta E_{DEL}$ includes the energy lowering due to electron transfer from the occupied ALMOs on one molecule to the virtual orbitals of another molecule, as well as the further energy change caused by induction (or repolarization) that accompanies such an occupied-virtual orbital mixing. Therefore, $\Delta E_{DEL}$ is further decomposed into the occupied-virtual forward- and back-donation components for each pair of molecules, plus the higher order (HO) induction energy:
\begin{eqnarray}
\label{eq:edeldec}
\Delta E_{DEL} &=& \sum_{x,y < x} \{ \Delta E_{x\rightarrow y} + \Delta E_{y\rightarrow x} \} + \Delta E_{HO},
\end{eqnarray}
$\Delta E_{HO}$ is generally small and cannot be naturally divided into two-body terms.
 
%

\textbf{Charge transfer analysis.} In addition to the energy decomposition, ALMOs have been used to measure the amount of electron density transferred between molecules~\cite{a:cta}. The intermolecular CT in such a charge transfer analysis (ALMO CTA), $\Delta Q_{DEL}$, is defined as the change in the electron density from the polarized $R_{POL}$ state to the fully converged state $R_{SCF}$. This definition is well-justified because, according to the Mulliken analysis, the ALMO constraint explicitly excludes CT between molecules, making $R_{POL}$ the ``zero-CT'' state with the lowest energy.

In agreement with ALMO EDA, $\Delta Q_{DEL}$ includes the charge transfer due to occupied-virtual mixing ($\Delta Q_{x\rightarrow y}$) and the accompanying higher order relaxation terms ($\Delta Q_{HO}$):
\begin{eqnarray}
\label{eq:qdeldec}
\Delta Q_{DEL} &=& \sum_{x,y < x} \{ \Delta Q_{x\rightarrow y} + \Delta Q_{y\rightarrow x} \} + \Delta Q_{HO}.
\end{eqnarray}
Thus, the intermolecular charge transfer terms in ALMO CTA have corresponding well-defined energies of stabilization calculated by ALMO EDA.

\textbf{Significant complementary occupied-virtual pairs.} In addition to quantifying the amount and energetics of intermolecular charge transfer, it is often useful to have a simple description of orbital interactions between molecules. The polarized canonical ALMOs used as a reference basis set in the decomposition analysis do not directly show which occupied-virtual orbital pairs are the most important in forming intermolecular bonds. That is, in general there are no occupied-virtual pairs in this basis set that can be neglected. However, orbital rotations within the occupied subset and within the virtual subset of a molecule leave $\Delta E_{x \rightarrow y}$ and $\Delta Q_{x \rightarrow y}$ unchanged. This freedom can be used to find new sets of orbitals for $x$ and $y$, in which charge transfer from $x$ to $y$ is described as \emph{each} occupied orbital on $x$ donating electrons to only \emph{one} (complementary) virtual orbital on $y$. Such orbitals are called complementary occupied-virtual pairs (COVPs)~\cite{a:cta}.

Construction of the COVPs greatly simplifies the picture of intermolecular orbital interactions since they form a ``chemist's basis set'' for a conceptual description of bonding in terms of just a few localized orbitals. As we demonstrate below, COVPs provide an alternative and somewhat unconventional view of hydrogen bonding in the water dimer. 

\textbf{Computational details and implementation.} ALMO EDA and CTA are implemented in both, the Q-Chem~\cite{a:qchem3} and CP2K~\cite{a:quickstep,www:cp2k} software packages. 
%
%
The variational optimization of the occupied ALMOs is performed using the locally projected SCF method~\cite{a:gia0,a:khal}. The definition of the ALMOs and the polarization energy relies on an underlying basis set that is partitioned amongst the fragments. Gaussian AO basis sets in both packages are ideal in this regard, and give well-defined polarization energies as long as there are no linear dependences. In the linearly dependent limit, where the basis functions on one fragment can exactly mimic functions on another fragment, this ceases to be the case~\cite{a:rvs}. This is not an issue for the AO basis sets used routinely in quantum chemistry and AIMD simulations~\cite{a:molopt}.

In addition to the occupied ALMOs, the locally projected SCF method yields a set of non-redundant linearly independent virtual ALMOs. After the SCF procedure is converged, the occupied subspace is projected out from the virtual ALMOs to ensure strong orthogonality of the subspaces. The energy lowering due to electron transfer from the occupied ALMOs on molecule $x$ to the virtual orbitals of molecule $y$ in Eq.~\ref{eq:edeldec} is a quasi-perturbative energy correction~\cite{a:khal,a:wzliang,a:wzliangeq} that can be expressed as:
\begin{equation} \label{eq:hatECT}
	\Delta E_{x\rightarrow y} = \sum_{i}^{o_x} \sum_{a}^{v_y} {F^{xi}}_{ya} {X^{ya}}_{xi},  
\end{equation}
where ${F^{xi}}_{ya}$ is the contravariant-covariant representation of the Fock operator build from the converged ALMOs and ${X^{ya}}_{xi}$ is the amplitude corresponding to the electron transfer (excitation) from the converged absolutely localized occupied orbital $i$ on fragment $x$ to the virtual orbital $a$ on fragment $y$. The variational nature of the polarized ALMOs guarantees that the energy term within a molecule is zero, $\Delta E_{x\rightarrow x} =0$.

The corresponding amount of charged transferred from $x$ to $y$ (Eq.~\ref{eq:qdeldec}) is expressed as~\cite{a:cta}:
\begin{equation} \label{eq:hatQCT}
	\Delta Q_{x\rightarrow y} = \sum_{i}^{o_x} \sum_{a}^{v_y} {X^{xi}}_{ya} {X^{ya}}_{xi} 
\end{equation}
The amplitudes ${X^{ya}}_{xi}$ are obtained by solving a quadratic equation~\cite{a:wzliangeq} using the preconditioned conjugate gradient method:
\begin{eqnarray} \label{eq:ampl}
{F^{ya}}_{xi} &+& \sum_{wb}^{V} {F^{ya}}_{wb} {X^{wb}}_{xi} - \sum_{yj}^{O} {X^{ya}}_{zj} {F^{zj}}_{xi} \nonumber \\
&-& \sum_{zj}^{O} \sum_{wb}^{V} {X^{ya}}_{zj} {F^{zj}}_{wb} {X^{wb}}_{xi} = 0
\end{eqnarray}
%
The total energy lowering due to the occupied-virtual mixing $\sum_{x,y} \Delta E_{x\rightarrow y}$ is equivalent to the result obtained by a single Fock matrix diagonalization~\cite{a:khal,a:wzliang}. The $\Delta E_{HO}$ term is introduced to recover the small difference between this energy and the fully converged SCF result.

It is important to note that special care must be taken to remove the basis set superposition error (BSSE) from the interaction energies and their components. The BSSE is not introduced when calculating frozen density and polarization energy contributions because constrained ALMO optimization prevents electrons on one molecule from borrowing the AOs of other molecules to compensate for the incompleteness of their own AOs. However, the BSSE enters the charge transfer terms since both the BSSE and charge transfer results from the same physical phenomenon of delocalization of fragment MOs. Therefore, these terms are inseparable from each other when finite Gaussian basis sets are used to describe fragments at finite spatial separation. It has been demonstrated that the BSSE decreases faster than charge transfer effects with increasing quality of the basis set~\cite{a:cteffects,a:nagatact,a:khal}. Therefore, the use of medium and large localized Gaussian basis sets (without linear dependencies) make the BSSE component of the interaction energy negligibly small. The BSSE associated with each forward- and back-donation term $\Delta E_{x\rightarrow y}$ can be corrected individually, as shown in Refs.~\onlinecite{a:theeda,a:cta}.

\textbf{Features of the ALMO decomposition methods.} ALMO EDA is conceptually similar to long-established decomposition methods, such as the Morokuma analysis~\cite{a:km}, but includes several important novel features described below.
\begin{itemize}
\item 
Unlike earlier decomposition methods~\cite{a:km,a:rvs,a:rvsX,a:csov,a:csov1,a:neda,a:nedaX,a:nedaDFT}, ALMO EDA and CTA treat the polarization term in a variationally optimal way. Therefore, CT effects (i.e. effects due to intermolecular electron delocalization) cannot be over- or underestimated.
\item
The CT term can be decomposed into forward-donation and back-bonding contributions for each pair of molecules in the systems.
\item
The ALMO charge transfer scale is such that all terms have well defined energetic effects. In contrast, population analysis methods include not only the true CT, but also large contaminating charge overlap effects~\cite{a:cta}. 
\item
COVPs constructed from canonical ALMOs provide a compact and chemically intuitive description of electron transfer between the molecules.
\item
The ALMO method in the CP2K package is currently the only decomposition scheme for condensed matter systems. CP2K relies on the mixed Gaussian and plane wave representation of electrons~\cite{a:gpw}, which makes it uniquely suited for performing ALMO calculations for periodic systems. In CP2K, the localized atom-centered Gaussian basis sets are used for the construction of ALMOs, whereas plane waves ensure the computational efficiency in large-scale calculations of the Hartree and XC potentials for periodic systems.
\end{itemize}

\section{Physical nature of hydrogen bonding in the water dimer}

As already alluded to, hydrogen bonding is central to all aqueous systems ranging from water nanoclusters and microsolvated ions to bulk water and solvated biomolecules~\cite{b:hbond,b:jeffrey,b:scheiner}. Despite numerous experimental and theoretical studies~\cite{a:darev,a:watdimrev,a:dyke,a:fellers,a:liu,a:saykally,a:compton,a:watdimcov}, the physical nature of hydrogen bonding is still under debate. One issue is the degree of covalency in the hydrogen bonding, which is determined by the extent of intermolecular electron delocalization or CT~\cite{a:compton,a:namebond,a:watbottom,a:watdimcov,a:watdimcov1}. Natural bond orbital (NBO)~\cite{a:nborev} and natural EDA~\cite{a:neda} suggests that CT is predominant~\cite{a:weinhold2006,a:nedaX,a:nedaDFT} because if CT is neglected, NBO analysis shows no binding at the water dimer equilibrium geometry. However, other earlier decomposition methods~\cite{a:km,a:csov,a:csov1,a:rvs,a:rvsX} estimated that CT contributes only around 20\% of the overall binding energy~\cite{a:rvs,a:rvsX,a:watcsov}. 

The extent of CT has practical significance for aqueous molecular dynamics simulations, where models based on purely electrostatic potentials (e.g. Coulomb perhaps with polarizability plus Lennard-Jones) seem to be very successful in reproducing many of the structural and thermodynamic properties of water~\cite{a:waterpotrev,a:jorgensen}. However, the failure of classical molecular dynamics simulations to reproduce the controversial ``chains and rings'' structure of liquid water, inferred from recent X-ray absorption and X-ray Raman scattering experiments~\cite{a:nilsson}, has generated questions about the reliability of existing water potentials~\cite{a:weinhold2006,a:nilsson,a:soper}. This fact, combined with the predominantly CT character of the hydrogen bonding suggested by NBO, has led to proposals to incorporate CT effects into empirical water potentials~\cite{a:weinhold2006}. 

In this section, we review the role of intermolecular CT effects in the simplest water cluster -- the water dimer -- uncovered with ALMO EDA and CTA~\cite{a:khalh2o}. Accurate separation of polarization effects from CT is essential to determine the amount of covalency in the hydrogen bonding. A variationally optimal treatment of polarization makes ALMO EDA and CTA ideal for this purpose.

\begin{figure}
\includegraphics[width=8cm]{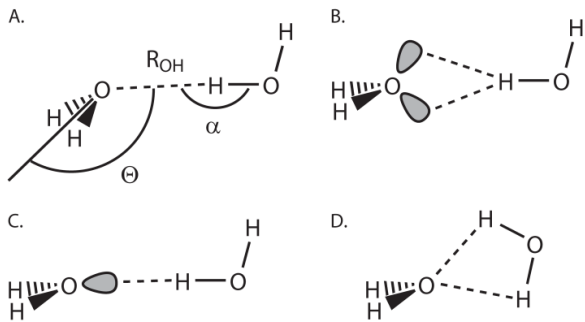}
\caption{\label{fig:dimer1} Water molecules in the water dimer.}
\end{figure}

\textbf{Energetic components of the hydrogen bond stabilization.} The relative position of the molecules in the water dimer with $C_{s}$ symmetry is described by three parameters shown in Figure~\ref{fig:dimer1}A. The structure of the dimer was optimized at the MP2/aug-cc-pVQZ level and is characterized by $\alpha = 172\degCC$, $\Theta = 127\degCC$, and $R_{OH} = 1.94\AA$. As mentioned above, the ALMO decomposition analysis is presently limited to single determinant wave functions. Therefore, ALMO EDA was applied to wave functions calculated at the Hartree-Fock and DFT level using a series of local density approximation (LDA), generalized gradient approximation (GGA), hybrid and meta-hybrid XC functionals (Table~\ref{tab:dimer3}).
   
The decomposition of the Hartree-Fock energy produces results similar to the earlier decomposition methods, but gives a somewhat larger charge transfer contribution (Table~\ref{tab:dimer3}). According to ALMO EDA, charge transfer accounts for 27\% of the total Hartree-Fock binding energy. When Kohn-Sham DFT is used instead of the Hartree-Fock method (Table~\ref{tab:dimer3}), all energy terms change because of modification of the exchange and addition of the correlation terms into the mean-field Hamiltonian. The delocalization effect becomes more pronounced for the density functional methods and in some cases the charge transfer term is more than 45\% of the overall binding energy. This observation is consistent with the tendency of modern density functionals to underestimate the HOMO-LUMO gap~\cite{a:perdewlevy,a:azadikuehne}, which in the water dimer case, manifests itself in a larger charge transfer energy. 

\begin{table*}
\caption{\label{tab:dimer3} The decomposition analysis results are reported for an LDA (SVWN~\cite{a:dirac,a:vwn}), three GGA (BP86~\cite{a:becke,a:p86}, PW91~\cite{a:pw91}), a hybrid (B3LYP~\cite{a:b3,a:lyp}) and an meta-hybrid XC functional (BMK~\cite{a:bmk}), as well as for the Hartree-Fock wave function. All terms are BSSE corrected and an aug-cc-pVQZ basis set is employed, throughout.}
\begin{ruledtabular}
\begin{tabular}{lcccccccccccc}
Scale & \multicolumn{6}{c}{$\Delta Q$, m\={e}} & \multicolumn{6}{c}{$\Delta E$, kJ/mol}\\
\hline
Functional & HF & BP86 & BMK & B3LYP & PW91 & SVWN & HF & BP86 & BMK & B3LYP & PW91 & SVWN\\
\hline
FRZ & \multicolumn{6}{c}{0.0} & -5.0 & -1.8 & -5.4 & -5.2 & -8.5 & -16.6 \\
POL & \multicolumn{6}{c}{0.0} & -6.1 & -7.0 & -6.6 & -6.5 & -5.2 & -7.9 \\
A$\rightarrow$D\footnotemark[1] & 0.1 & 0.2 & 0.1 & 0.1 & 0.2 & 0.2 & -0.2 & -0.3 & -0.2 & -0.3 & -0.4 & -0.3 \\
D$\rightarrow$A\footnotemark[1] &0.8 & 4.0 & 1.4 & 2.8 & 4.7 & 4.3 & -3.2 & -8.1 & -5.1 & -6.6 & -8.1 & -8.1 \\
Rem. CT\footnotemark[2] & 0.2 & -1.1 & 0.3 & -0.4 & -1.8 & -1.4 & -0.6 & 0.1 & -0.7 & -0.3 & 0.1 & 0.1 \\
\textbf{TOT}\footnotemark[3] & \textbf{1.1} & \textbf{3.1} & \textbf{1.8} & \textbf{2.5} & \textbf{3.1} & \textbf{3.1} & \textbf{-15.1} & \textbf{-17.1} & \textbf{-17.9} & \textbf{-18.9} & \textbf{-22.1} & \textbf{-32.8} \\
BSSE & 0.0 & 0.0 & 0.1 & 0.0 & 0.0 & 0.0 & 0.1 & 0.2 & 0.6 & 0.1 & 0.2 & 0.2 \\
COVP$_{1}$\footnotemark[4] & 96 & 97 & 83 & 96 & 95 & 96 & 94 & 95 & 85 & 93 & 90 & 93 \\
\end{tabular}
\end{ruledtabular}
\footnotetext[1]{D -- electron-donor (proton-acceptor), A -- electron-acceptor (proton-donor).}
\footnotetext[2]{Remaining CT includes intramolecular terms as well as the higher order relaxation term.}
\footnotetext[3]{BSSE corrected MP2/aug-cc-pVQZ total interaction energy is -20.6~kJ/mol.}
\footnotetext[4]{Contribution of COVP$_{1}$ is given as percent of $D\rightarrow A$.}
\end{table*}

It is clear from Table~\ref{tab:dimer3} that all three energy components (frozen density, polarization, and charge-transfer) are important for the energetic stabilization of the dimer at its equilibrium geometry. We, therefore, conclude that the NBO approach significantly overestimates CT due to the non-variational treatment of the reference ``zero-CT'' electronic state. Our results show that CT contributes around one third of the overall binding energy in the complete basis set limit, of which approximately 95\% is from the proton acceptor to the proton donor. The same effect is observed on the charge scale, indicating a direct correspondence between electron redistribution and the energy of CT interactions.

\begin{figure}
\includegraphics[width=8.0cm]{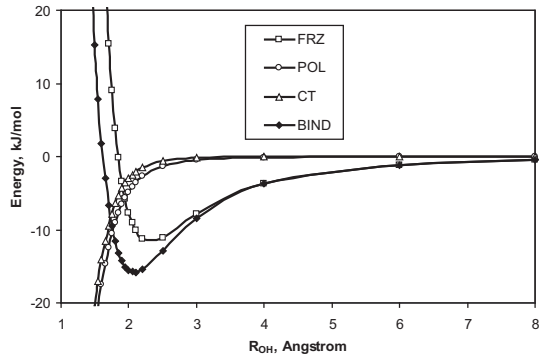}
\caption{\label{fig:F3e} Dependence of the energy components on the distance between the water molecules in the water dimer at the HF/aug-cc-pVQZ level of theory.}
\end{figure}

The relative contribution of the energy terms varies strongly with the position of the molecules in the water dimer. Figure \ref{fig:F3e} shows the dependence of the Hartree-Fock BSSE corrected energy and its ALMO decomposition on the distance between the water molecules ($R_{OH}$ is varied and all other internal coordinates remain fixed at their MP2/aug-cc-pVQZ values). The frozen density component increases significantly and becomes repulsive as the molecules get closer. At the same time, stabilization due to polarization and charge transfer increases upon the closer contact, but not strongly enough to compensate for the electron density repulsion. The stabilizing contribution of charge transfer and polarization decrease rapidly with the increase of the intermolecular distance and, at $R_{OH} > 3$\AA, the interaction energy can be accurately approximated by the frozen density term alone. 


Table~\ref{tab:dimer3} shows that the relative contribution of the three energy components somewhat depends on the XC functional chosen, just as the total binding energies do. Such a wide spread of the DFT results indicates that a better description of the intermolecular correlation energy is necessary for the water dimer. At the same time, the main \emph{qualitative} description of hydrogen bonding remains the same for all commonly used XC functionals. Therefore, we further discuss the results obtained with the B3LYP XC functional. This functional most closely reproduces more accurate MP2 water dimer binding energies of the various XC functionals that we tried (Table~\ref{tab:dimer3}). A detailed discussion of the performance of popular density functionals for describing HB between water molecules can be found in Refs.~\onlinecite{a:xugoddard,a:app_almo_4,a:app_almo_11}.

The results of the B3LYP ALMO decomposition analysis are presented in Table~\ref{tab:dimer1}, which shows that the energy and charge components rapidly converge as the basis set becomes locally complete, indicating the high stability of the ALMO decomposition. All CT terms presented here are BSSE corrected. The BSSE is presented in Table~\ref{tab:dimer1} to show the degree of basis set completeness. The very small values of BSSE suggests that the aug-cc-pVQZ and aug-cc-pV5Z basis sets are effectively complete for the present purpose.

\begin{table*}
\caption{\label{tab:dimer1} ALMO CTA and EDA results for the water dimer (B3LYP/aug-cc-pVYZ). All terms are BSSE corrected.}
\begin{ruledtabular}
\begin{tabular}{lcccccccc}
Scale & \multicolumn{4}{c}{$\Delta Q$, m\={e}} & \multicolumn{4}{c}{$\Delta E$, kJ/mol} \tabularnewline
\hline
Y & D & T & Q & 5 & D & T & Q & 5 \tabularnewline
\hline
FRZ & \multicolumn{4}{c}{0.0} & -5.5 & -5.4 & -5.2 & -5.1\\
POL & \multicolumn{4}{c}{0.0} & -4.5 & -6.1 & -6.5 & -7.1\\
A$\rightarrow$D\footnotemark[1] & 0.1 & 0.2 & 0.1 & 0.1 & -0.2 & -0.4 & -0.3 & -0.2\\
D$\rightarrow$A\footnotemark[1] & 3.7 & 2.4 & 2.8 & 2.7 & -7.9 & -6.6 & -6.6 & -6.3\\
Rem. CT\footnotemark[2] & 0.3 & 0.2 & -0.4 & -0.5 & -0.4 & -0.3 & -0.3 & -0.3\\
\textbf{TOT}\footnotemark[3] & \textbf{4.0} & \textbf{2.7} & \textbf{2.5} & \textbf{2.3} & \textbf{-18.5} & \textbf{-18.8} & \textbf{-18.9} & \textbf{-18.9} \\
BSSE & 0.3 & 0.1 & 0.0 & 0.0 & 1.0 & 0.2 & 0.1 & 0.0\\
COVP$_{1}$\footnotemark[4] & 95 & 97 & 96 & 97 & 90 & 97 & 93 & 96\\
\end{tabular}
\end{ruledtabular}
\footnotetext[1]{D -- electron-donor (proton-acceptor), A -- electron-acceptor (proton-donor).}
\footnotetext[2]{Remaining CT includes intramolecular terms as well as the higher order relaxation term.}
\footnotetext[3]{MP2 interaction energies are -18.3, -19.8, and -20.6~kJ/mol for Y=D, T, Q, respectively.}
\footnotetext[4]{Contribution of COVP$_{1}$ is given as percent of $D\rightarrow A$.}
\end{table*}

\textbf{The role of electron transfer in hydrogen bonding.} The total electron density transfer calculated with ALMO CTA is just a few milli-electrons (2.3m\={e} at the B3LYP/aug-cc-pV5Z level). This result is an order of magnitude smaller than the CT calculated with the Mulliken, L\"owdin and natural population analysis (PA) methods (Table~\ref{tab:dimer2})~\cite{a:mulliken2,a:nonproblem,a:npa}. This discrepancy arises largely from the different meaning assigned to CT in ALMO CTA and PA techniques. ALMO CTA measures CT as the degree of electron relaxation from the optimal polarized (pre-CT) state to the delocalized state. By contrast, PA methods include not only the true CT, but also the separate and, in this case, larger effect of partitioning the charge distribution of the polarized pre-CT state (for a detailed comparison see Ref.~\onlinecite{a:cta}). Thus, the key advantage of the ALMO CTA approach is that it shows the electron transfer associated with an energy lowering due to dative interactions: just a few milli-electrons. 

\begin{table}
\caption{\label{tab:dimer2} Charge (m\={e}) of the electron-acceptor molecule in the dimer (B3LYP/aug-cc-pVXZ). All charges are BSSE corrected.}
\begin{ruledtabular}
\begin{tabular}{lcccc}
X & D & T & Q & 5 \\
\hline
Mulliken PA & -27.5 & -18.8 & -22.2 & -17.0 \\
L\"owdin PA & -24.0 & -24.0 & -21.2 & -17.8 \\
Natural PA & -18.3 & -16.5 & -17.1 & -- \\
\end{tabular}
\end{ruledtabular}
\end{table}

It may seem remarkable that so little CT can stabilize the HB by 6.5~kJ/mol (equivalent to 32~eV per electron), but this estimate is consistent with simple estimates from perturbation theory. The CT energy is a second order correction to the energy of the polarized system, and is proportional to $\frac{F^{2}_{ad}}{(\epsilon_a-\epsilon_d)}$, where $F_{ad}$ is the CT energy coupling between donating orbital $d$ and accepting orbital $a$, while $\epsilon_i$ is the energy of orbital $i$. The $\Delta Q$ term, however, is proportional to $\frac{F^{2}_{ad}}{(\epsilon_a-\epsilon_d)^{2}}$. Therefore, the CT energy per electron is related to the energy gap between donating and accepting orbitals $\epsilon_a-\epsilon_d$. The B3LYP/aug-cc-pV5Z energy gap between the most important donating and accepting orbitals in the water dimer lies between 10 and 40 eV (virtual orbitals in such a big basis set practically form a continuum of states). Thus, a value of 32~eV for the effective $d-a$ gap for CT between water molecules in the dimer at the equilibrium geometry is reasonable.

\begin{figure}
\includegraphics[width=8cm]{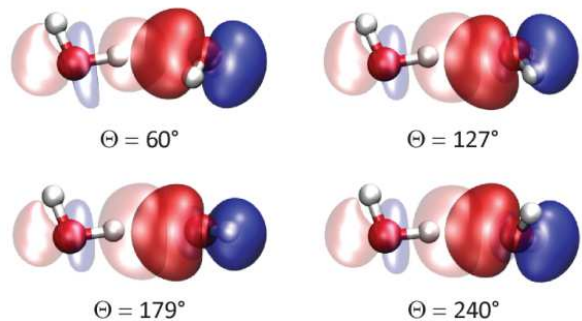}
\caption{\label{fig:dimer2} Dependence of the shape of the most significant COVP on the relative orientation of the water molecules in the dimer. All figures show orbitals calculated at the B3LYP/aug-cc-pVQZ level of theory. Isosurface value of 0.05 a.u. Occupied orbitals are represented with saturated colours. Faint colours represent complementary virtual orbitals.}
\end{figure}

\textbf{Orbital representation of donor-acceptor interactions.} COVPs let us visualize CT effects and provide additional insight into the nature of hydrogen bonding. In the water dimer, only one COVP is significant, and recovers 96-97\% of the overall transfer an from proton acceptor to a proton donor on the energy and charge scales (Table~\ref{tab:dimer1}). The remaining CT in this direction can be attributed to the four remaining COVPs, none of which exceeds 3\% of the overall transfer. The shapes of the occupied and the virtual orbitals of the main COVP are shown in Figure~\ref{fig:dimer2} ($\Theta$~=~127\degCC). The virtual orbital resembles the O-H anti-bonding orbital, $\sigma^{*}_{OH}$, of the electron accepting molecule. This is consistent with oxygen atom lone pairs donating electron density to the anti-bonding orbitals of the other molecule. However, the shape of the donating orbital is somewhat unexpected because it does not resemble an $sp^{3}$-hybridised lone pair.

\begin{figure}
\includegraphics[width=8cm]{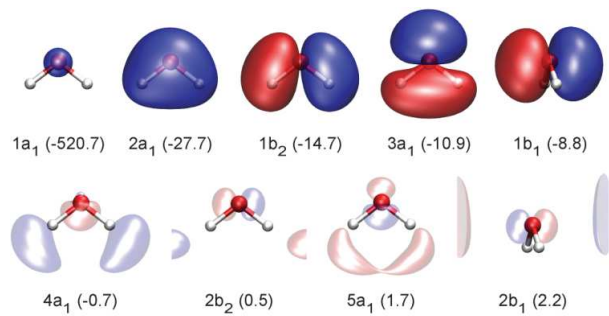}
\caption{\label{fig:dimer3} Symmetry of the five occupied and the lowest four virtual canonical MOs of water molecule. See the caption of Figure~\ref{fig:dimer2} for full description.}
\end{figure}

Commonly, $sp^{3}$-hybrids play an important role in predicting geometries of gas-phase water molecules and direction of HBs in ice phases. Furthermore, $sp^{3}$-hybridised lone pairs on the O atom have become a commonly accepted way to visualize the electronic structure of water molecules and are the basis for the five-point molecular models (e.g. TIP5P) widely used in classical molecular dynamics simulations~\cite{a:jorgensen}. However, the occupied orbitals are not unique, and can, therefore, be fixed by criteria that include well-defined ionization energy (giving canonical orbitals, Figure~\ref{fig:dimer3}), maximal localization (giving $sp^{3}$ lone pairs), or the most compact representation of CT effects (giving the COVP shown in Figure~\ref{fig:dimer2}). The COVP is best suited for studying donor-acceptor interactions, and the form of the optimal donor and acceptor orbitals can be understood as a compromise between high energy and good interaction with the acceptor. In this regard, the optimal acceptor orbital of Figure~\ref{fig:dimer2} bears almost no resemblance to the low-lying canonical virtual orbitals of Figure~\ref{fig:dimer3}, consistent with the effective $d-a$ gap being far larger than the HOMO-LUMO gap. The occupied (donating) orbital is mostly a linear combination of $3a_1$ and $1b_1$ canonical orbitals (Figure~\ref{fig:dimer3}) that are the two highest lying orbitals of the electron donating water molecule. A small CT contribution from the $2a_1$ canonical orbital of the donating molecule is reasonable given that its low energy makes it a poor donor. This simple argument explains the form of the donating orbital in the water dimer complex (Figure~\ref{fig:dimer2}, $\Theta = 127\degCC$).

\begin{figure}
\includegraphics[width=8cm]{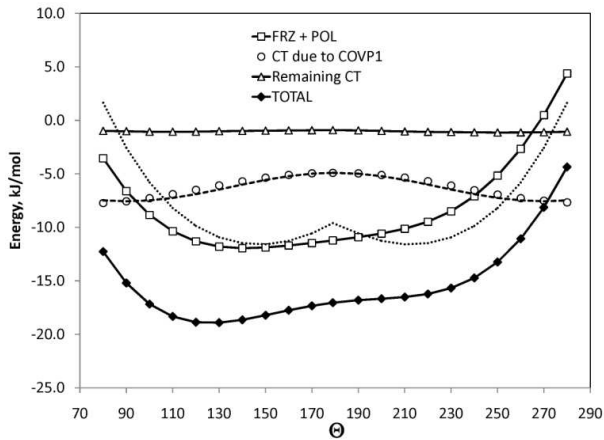}
\caption{\label{fig:dimer4} Dependence of the B3LYP/aug-cc-pVQZ energy components on the relative orientation of water molecules in the dimer. Dashed line represents CT calculated according to Eq.~\ref{eq:rotct}. Dotted line represents FRZ+POL interactions calculated according to Eq.~\ref{eq:rotfrzpol}.}
\end{figure}

Further support for this interpretation comes from Figure~\ref{fig:dimer4}, which shows the dependence of the B3LYP/aug-cc-pVQZ BSSE corrected energy and its ALMO decomposition on the orientation of the water molecules ($\Theta$ is varied and all other internal coordinates remain fixed at their MP2/aug-cc-pVQZ values). The CT energy does not maximize around tetrahedral coordination ($\Theta = 127\degCC$). The energy lowering due to the most significant COVP changes remarkably little from -4.9~kJ/mol for $\Theta = 180\degCC$ to -7.7~kJ/mol for $\Theta = 50\degCC$. From Figure~\ref{fig:dimer2}, the donor orbital does not rotate with the water molecule but stays directed towards the electron accepting molecule, unlike an $sp^{3}$ lone pair. The principal donor orbital thus changes with rotation to optimize the coupling with the complementary 
$\sigma^{*}_{OH}$ virtual acceptor orbital, thereby explaining the weak dependence of the CT energy on $\Theta$.

It is interesting to consider HBs that involve bifurcated interactions. $\Theta = 179\degCC$ corresponds to an OH bond interacting with two $sp^{3}$ lone pairs, which represent a bifurcated proton donor in the traditional picture (see the cartoon in Figure~\ref{fig:dimer1}B). However, the CT contribution to H-bonding still involves only one donating orbital, as shown in Figure~\ref{fig:dimer2} and the cartoon in Figure~\ref{fig:dimer1}C. The reduction in CT energy reflects at $\Theta = 179\degCC$ a greater contribution of the lower energy $3a_1$ orbital and a decreased contribution of the highest occupied $1b_1$ orbital of the donor molecule. In fact, the CT energy dependence on $\Theta$ can be well represented as a linear combination of CT from these two orbitals by a simple equation (the dashed line in Figure~\ref{fig:dimer4}):
\begin{equation} \label{eq:rotct}
\Delta E_{D\rightarrow A}(\Theta) = \Delta E_{1b_1} \sin^2(\Theta) + \Delta E_{3a_1} \cos^2(\Theta)
\end{equation}

The shape of the FRZ curve can be explained in purely electrostatic terms. The dotted line in Figure~\ref{fig:dimer4} represents the interaction energy of point charges placed at the position of the nuclei in the dimer ($-1.0$\={e} and $+0.5$\={e} charges replace O and H atoms correspondingly) and is given by equation:
\begin{equation} \label{eq:rotfrzpol}
\Delta E_{FRZ+POL}(\Theta) = \sum_{i\in D} \sum_{j\in A} \frac{q_i q_j}{r_{ij}} + 29.2\text{ kJ mol}^{-1}
\end{equation}

The constant in the equation is included to capture effects that are essentially independent of $\Theta$ such as polarization and exchange. Therefore, the position of the minimum on the total energy curve at $\Theta = 127\degCC$ is determined by combination of both electrostatic and charge transfer interactions. 

\begin{figure}
\includegraphics[width=8cm]{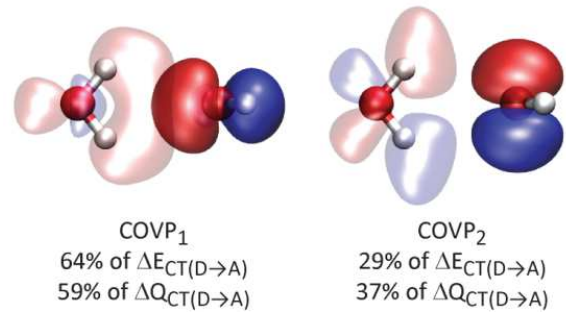}
\caption{\label{fig:dimer5} Bifurcated hydrogen bonding in the water dimer. EDA terms, kJ/mol: $\Delta E_{FRZ} = -8.2$, $\Delta E_{POL} = -1.7$, $\Delta E_{D\rightarrow A} = -0.9$, $\Delta E_{A\rightarrow D} = -0.2$, $\Delta E_{TOT} = -11.0$. CTA terms, m\={e}: $\Delta Q_{D\rightarrow A} = 0.4$, $\Delta Q_{A\rightarrow D} = 0.1$, $\Delta Q_{TOT} = 0.5$. See the caption of Figure~\ref{fig:dimer2} for full description.}
\end{figure}

In the case of a bifurcated proton acceptor (Figure~\ref{fig:dimer1}D), the description of HB changes qualitatively. The CT term becomes very small due to poor interactions and has significant contributions from two COVPs -- the symmetric ($3a_1$ canonical orbital) and anti-symmetric ($1b_1$ canonical orbital) (Figures~\ref{fig:dimer3} and~\ref{fig:dimer5}). This reflects the availability of two acceptor anti-bonding $\sigma^{*}_{OH}$ (symmetric $4a_1$ and anti-symmetric $2b_2$) orbitals in the vicinity of the electron donating molecule. It is certain that the donating orbitals will change their shape and orientation in larger water clusters and bulk liquid water according to the local environment.

\textbf{Summary.} ALMO energy decomposition and charge transfer analysis is a promising method to study hydrogen bonding between water molecules using density functional theory. It shows that although electron delocalization effects play an important role in hydrogen bonding they are not solely responsible for the energetic stabilization of the water dimer. The contributions of frozen density interactions and polarization are not less significant than that of CT, unlike some earlier work~\cite{a:weinhold2006,a:nedaX,a:nedaDFT}. ALMO CTA demonstrates that the amount of intermolecular CT is in the order of few milli-electrons, which is much smaller than has been inferred from population analyses. It also shows that CT is fairly insensitive to intermolecular rotation of the water molecules. This helps to account for the success of empirical potentials that do not include charge transfer explicitly. Furthermore, COVPs provide a new view of the electron donating orbital in the water dimer. Unlike rigid $sp^3$ lone pairs, the COVP donor changes its orientation according to the relative positions of the two molecules. A single $p$-like lone pair is usually the dominant donor, although at the geometry of a bifurcated HB, the CT contribution becomes small and two donor orbitals contribute. 

\textbf{Further applications of ALMO EDA to water clusters.} ALMO EDA has been applied to investigate the importance of charge transfer effects for the vibrational spectrum of the water dimer~\cite{a:app_almo_2}. Comparing the vibrational spectra calculated with and without the charge-transfer shows that electron delocalization has a very large effect on the vibrational frequency and intensity associated with the stretch of the donated OH bond, while its effect on the other vibrational modes is small~\cite{a:app_almo_2}. Further applications of ALMO EDA to water clusters include a recent study of HB in trimer, tetramer, pentamer, 13-er, and 17-er~\cite{a:app_almo_11}. The decomposition of the two- three- and higher-body interaction energies into the frozen-density, polarization and charge-transfer components has revealed several interesting trends that provide additional physical justification for the standard practice of not explicitly including charge transfer into water force fields. ALMO EDA has also been used, in combination with other decomposition schemes, to gain insight into the performance of several popular density functionals for describing interactions between water molecules~\cite{a:app_almo_4}.

\section{Liquid water}

\subsection{Structural and dynamical properties of liquid water}

Despite the ongoing development of new simulation techniques, an accurate modelling of liquid water still represents a major challenge. The accuracy of AIMD simulations has often to be sacrificed to reduce their computational burden. Therefore, various physical effects such as van der Waals interactions between molecules and the quantum behaviour of nuclei have to been reproduced only approximately or even completely neglected. In addition, the high computational cost of AIMD imposes severe restrictions on the size of a model system and time length of the simulations thus introducing additional finite-size errors and statistical uncertainties into the properties calculated~\cite{a:asthagiri, a:CP2Kwater1, a:fernandez-serra, a:CP2Kwater2, a:sit, a:mantz, a:lee1, a:lee2, a:BanyaiSebastiani}.

In this section, we assess the magnitude of errors coming from physical approximations, finite-size effects and insufficient sampling by performing large-scale simulations, which exploit the computationally efficiency second-generation CPMD method~\cite{a:2ndcpmd, a:kuhnewater}.


\textbf{Computational details.} The largest simulated system consisted of 128 light water molecules in a periodic cubic box of length $L = 15.6627$~{\AA}, which corresponds to a density that differs by only $0.3$~\% from the experimental value. All simulations were performed at 300~K within the canonical NVT ensemble; the Langevin equation of motion was integrated using the algorithm of Ricci and Ciccotti~\cite{a:RicciCiccotti}. The discretized integration time step $\Delta t$ was set to 0.5~fs, while $\gamma_{D} = 8.65 \times 10^{-5}$~fs$^{-1}$. The simultaneous propagation of the electronic degrees of freedom proceeded with $K=7$, which yields a time reversibility of $\mathcal{O}(\Delta t^{12})$. At each MD step the corrector was applied only once, which implies just one preconditioned gradient calculation. The deviation from the BO surface, as measured by the preconditioned mean gradient deviation was $10^{-5} a.u.$, which is only slightly larger than typical values used in fully converged BOMD simulations. The Brillouin zone was sampled at the $\Gamma$-point only and, unless stated otherwise, the PBE XC functional has been employed~\cite{a:pbe}. Separable norm-conserving pseudopotentials were used to describe the interactions between the valence electrons and the ionic cores~\cite{a:gth, a:hgh, a:krackpp}.

Long and well-equilibrated trajectories were necessary to obtain an accurate sampling. This requirement was made even more stringent by the strong  dependence of the translational self-diffusion coefficient on temperature and, in the case of PBE water, on the expected low diffusivity at room temperature~\cite{a:sit, a:grossman}. Therefore, in each run, the system was equilibrated for $25$~ps and the statistics were accumulated for the successive $250$~ps. Finite-size effects are studied by comparing the results of the largest system with equally long runs on $64$ and $32$ water molecule. Two shorter $25$~ps BOMD reference calculations with 128 molecules were carried out using either Newtonian or Langevin dynamics to assess the accuracy of the simulations. The settings for both runs were identical and started from the same well-equilibrated configuration. The influence of the XC functional on the properties calculated was investigated in a series of additional runs using a variety of different semilocal XC functionals~\cite{a:RPBE, a:revPBE, a:becke, a:OPTX, a:lyp}. The statistics in each of these reference runs were accumulated for at least $30$~ps after an equilibration of $20$~ps, resulting in more than $1$~ns of AIMD simulations.

All the simulations were performed using the CP2K/\textsc{Quickstep} code~\cite{www:cp2k, a:quickstep}, which relies on the mixed Gaussian and plane wave representation of the electronic degrees of freedom~\cite{a:gpw}. In this approach, the Kohn-Sham (KS) orbitals are expanded in terms of Gaussian orbitals, while a plane wave basis is used for the electron density. Such a dual basis approach combined with advanced multigrid, sparse matrix and screening techniques enables one to achieve an efficient linear-scaling evaluation of the KS matrix. Efforts towards a full inear scaling algorithm are underway~\cite{a:ceriottikuehne1, a:ceriottikuehne2, a:richterskuehne}. Here, the orbitals were represented by an accurate triple-$\zeta$ basis set with two set of polarization functions (TZV2P)~\cite{a:molopt}, while a density cutoff of $320$~Ry was used for the charge density.

\begin{figure}
  \includegraphics[width=8cm]{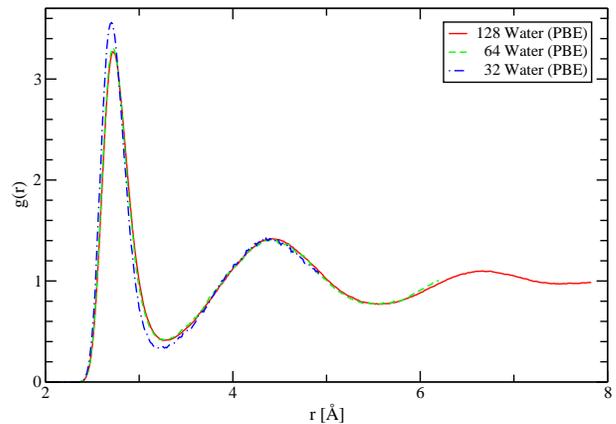}
  \caption{\label{fig:water1} Comparison of $g_{OO}(r)$ as obtained from second-generation CPMD simulations with 32, 64 and 128 water molecules.}
\end{figure}

\textbf{Structural properties.} The influence of finite-size effects on the oxygen-oxygen radial distribution function (RDF) $g_{OO}(r)$ is shown in Figure~\ref{fig:water1}. It can be seen that the RDF is quickly converging with respect to system size: the $g_{OO}(r)$ for the 64-molecule system coincides within statistical uncertainties with the result of a larger 128-molecule simulation.

\begin{figure*}
  \includegraphics[width=16cm]{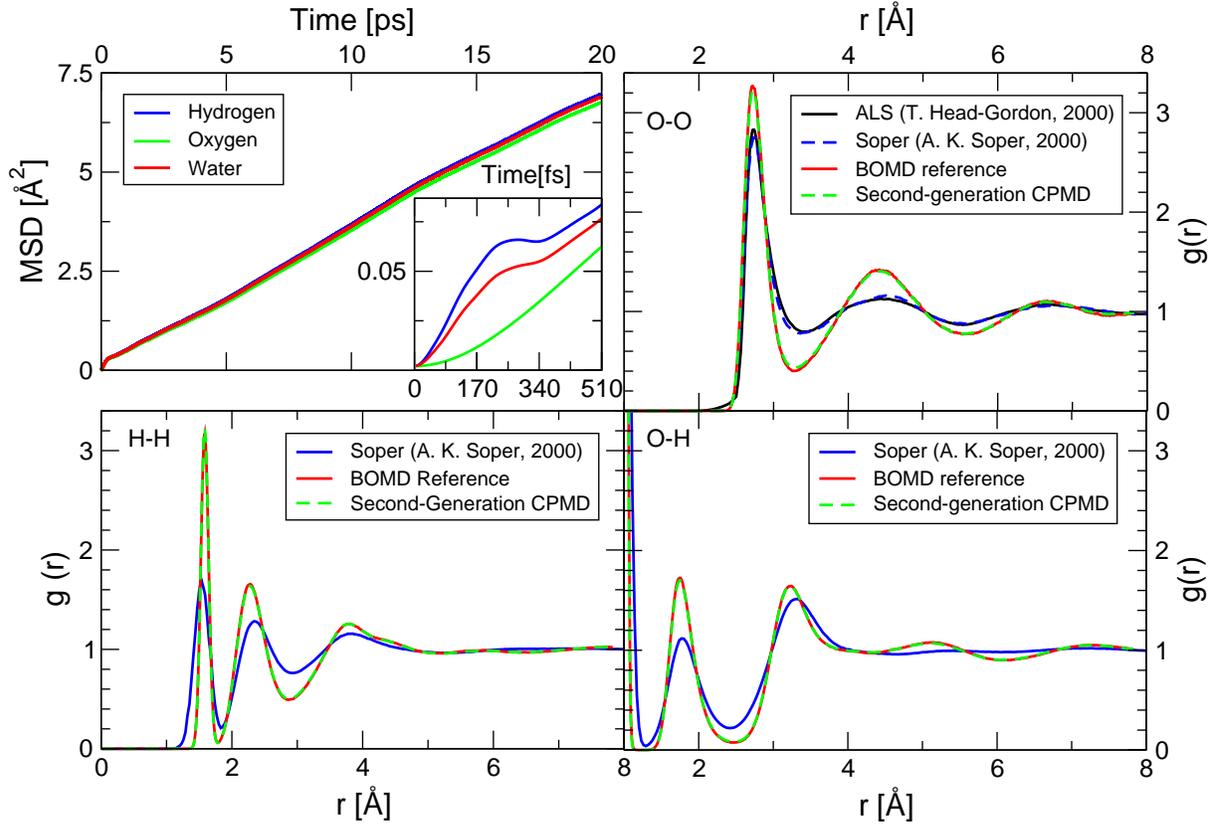}
  \caption{\label{fig:water2} Partial RDFs of liquid water at ambient conditions and its mean square displacement (top left panel). In the corresponding inset the onset of the cage effect can be observed at $\sim250$~fs followed by diffusion, which is in excellent agreement with Gallo et al.~\cite{a:gallo}.
  }
\end{figure*}

The accuracy of the second-generation CPMD can be readily established by comparing the calculated RDFs to the BOMD reference (Figure~\ref{fig:water2}). While the agreement between these two methods is excellent, they both predict water to be overstructured, which can be seen by comparing the present RDFs with the ones obtained from recent neutron diffraction~\cite{a:soper-water-xray} and X-ray scattering~\cite{a:thg-xray-water-2003} experiments (Figure~\ref{fig:water2}). The most pronounced disagreement between the calculated and experimentally derived RDFs is the case of $g_{OH}(r)$, for which the relative heights of the first two intramolecular peaks is reversed. However, the inclusion of nuclear quantum effects~\cite{a:KleinParrinello, a:MorroneSebastianiCar, a:MorroneCar, a:habershon}, as well as London dispersion forces~\cite{a:SchmidtVdW, a:kuehnewatersurface, a:artachoVdW, a:rothlisbergerVdW, a:tuckermanVdW} is expected to improve agreement with experiment. Using either artificially increased temperatures~\cite{a:morrone, a:CP2Kwater2, a:sit, a:mantz, a:lee1, a:lee2} or different XC functionals~\cite{a:asthagiri, a:fernandez-serra, a:CP2Kwater1} may also lead to a better agreement with the experimental data.

\begin{figure}
  \includegraphics[width=8cm]{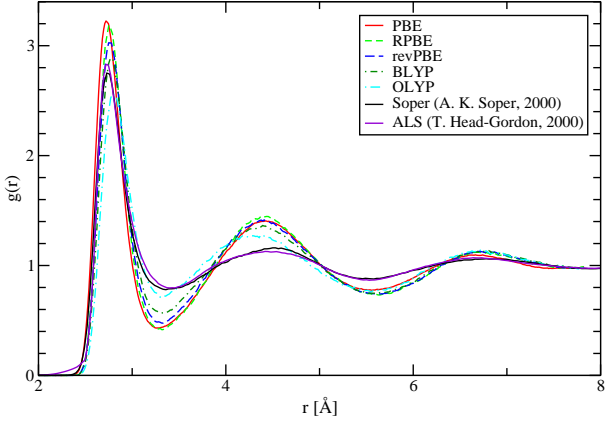}
  \caption{\label{fig:water3} Comparison of $g_{OO}(r)$ obtained from neutron diffraction, X-ray scattering and second-generation Car-Parrinello simulations using a variety of different XC functionals.}
\end{figure}

After verifying that the RDFs are converged with respect to the system size (i.e. finite-size effects are negligible) a series of simulations with commonly used XC functionals were performed. Figure~\ref{fig:water3} and Table~\ref{StaticWater} show that the shapes of the RDFs depend strongly on the particular XC functional, whereas the coordination number for a water molecule as obtained by integrating the RDF curve up to the first minimum remains very close to four for all XC functionals.

Obtaining a meaningful coordination number has important implications for the debate around the structure and symmetry of the HB network in liquid water. The debate stems from an interpretation of the X-ray spectra of water as evidence for a large fraction of molecules ($\sim$80\%) with broken HBs and a two-fold coordination. To check the validity of this interpretation, their spectroscopically derived HB definition~\cite{a:nilsson} was used to calculate the average number of bonds formed by a water molecule. The results presented in Table~\ref{WaterHBond} show that the average coordination number depends somewhat on the XC functional and varies between 3.1 (OLYP) to 3.6 (PBE). Given the fact that these results do not change qualitatively upon altering the HB definition~\cite{a:luzar, a:aimd3, a:kuehnewatersurface}, our AIMD simulations support a conventional nearly-tetrahedral view on the structure of liquid water. In the next section, we will analyze the symmetry and average number of HBs using their electronic signatures in addition to the simple geometric criteria utilised here.


\begin{table}[b]
  \caption{The position and height of the first maximum and minimum of $g_{\text{OO}}(r)$ and the coordination number $N_{c}$ as calculated by integrating $g_{\text{OO}}(r)$ up to the first minimum.}
  \begin{tabular}{lccccc}
    \hline \hline
    XC & $g_{\text{OO}}^{\text{max}}(r)$ & $r_{\text{OO}}^{\text{max}}$ & $g_{\text{OO}}^{\text{min}}(r)$ & $r_{\text{OO}}^{\text{min}}$ & $N_{c}$ \\
    \hline
    PBE~\cite{a:pbe} & 3.25 & 2.73 & 0.44 & 3.28 & 4.04 \\
    RPBE~\cite{a:RPBE} & 3.19 & 2.75 & 0.42 & 3.32 & 4.03 \\
    revPBE~\cite{a:revPBE} & 3.01 & 2.77 & 0.50 & 3.31 & 4.05 \\
    BLYP~\cite{a:becke, a:lyp} & 2.92 & 2.79 & 0.57 & 3.33 & 4.09 \\
    OLYP~\cite{a:OPTX, a:lyp} & 2.57 & 2.79 & 0.71 & 3.30 & 3.90 \\
    Soper~\cite{a:soper-water-xray} & 2.75 & 2.73 & 0.78 & 3.36 & - \\
    ALS~\cite{a:thg-xray-water-2003} & 2.83 & 2.73 & 0.80 & 3.4 & 4.7 \\
    \hline \hline
  \end{tabular}
  \label{StaticWater}
\end{table}

\begin{table}[b]
  \caption{The relative occurrence of double donor (DD), single donor (SD), no donor (ND) water molecules, percentage of donated and free HBs, as well as the average number of HBs formed by a water molecule. The results were obtained from AIMD simulations using several semi-local XC functionals.}
  \begin{tabular}{lrrrrrr}
    \hline \hline
     & PBE & RPBE & revPBE & BLYP & OLYP \\
    \hline
    DD & 82.8 \% & 81.4 \% & 76.8 \% & 72.9 \% & 59.0 \% \\
    SD & 16.6 \% & 17.8 \% & 22.0 \% & 25.4 \% & 36.3 \% \\
    ND & 0.7 \% & 0.8 \% & 1.2 \% & 1.7 \% & 4.7 \% \\
    \hline
    donated HB's & 91.0 \% & 90.3 \% & 87.8 \% & 85.6 \% & 77.1 \% \\
    free HB's & 9.0 \% & 9.7 \% & 12.2 \% & 14.4 \% & 22.9 \% \\
    \hline
    mean HB's & 3.642  & 3.613  & 3.513  & 3.423  & 3.085  \\
    \hline \hline
  \end{tabular}
  \label{WaterHBond}
\end{table}

\textbf{Dynamical properties.} Compared to the structural properties, the translational diffusion constant is known to exhibit stronger dependence on the size of the simulation box. The former is due to the fact that a diffusing particle entails a hydrodynamic flow, which decays rather slowly as 1/r. In a periodically repeated system this leads to an interference between a single particle and its periodic images. Analysing this effect, D\"unweg and Kremer~\cite{a:thomas63} have derived the following relation:
\begin{eqnarray}
   D(\infty) = D(L) + \frac{k_BT \zeta}{6 \pi \eta L}, \label{DunwegKremer}
\end{eqnarray}
where $D(L)$ is the translational diffusion coefficient calculated for a system with the length of the simulation cell $L$, $\eta$ is the translational shear viscosity, whereas the constant $\zeta$ is equal to 2.837.

\begin{figure}
  \includegraphics[width=8cm]{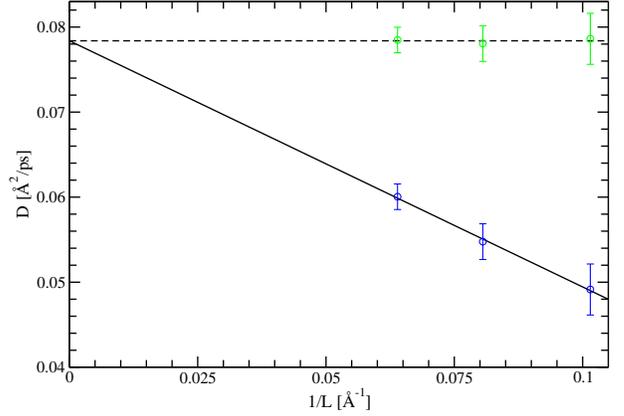}
  \caption{\label{fig:water4} The translational diffusion coefficient as a function of system size computed by second-generation Car-Parrinello simulations using the PBE XC functional. The solid line is the linear extrapolation to $D(\infty)$ whereas the dotted line is the mean $D(\infty)$ obtained from Eq.~\ref{DunwegKremer}.}
\end{figure}

Figure~\ref{fig:water4} shows that the 1/L dependence predicted by Eq.~\ref{DunwegKremer} holds rather well for the present AIMD simulations, and thus, can be used to determine extrapolated values for the translational diffusion coefficient $D(\infty)=0.789 \times 10^{-5}\,\text{cm}^{2}/\text{s}$ and the shear viscosity coefficient $\eta = 21.22 \times 10^{-4}\,\text{Pa} \cdot \text{s}$. Comparison with the experimental values $D=2.395 \times 10^{-5}\,\text{cm}^{2}/\text{s}$~\cite{a:hardy} and $\eta=8.92 \times 10^{-4}\,\text{Pa} \cdot \text{s}$~\cite{a:HarrisWoolf} confirms that liquid water as obtained from PBE AIMD simulations is less fluid than real water. However, taking into account that nuclear quantum effects were completely neglected in the simulations, it is possible to conclude that the PBE XC functional provides a rather good representation of liquid water, much better than generally appreciated. 

Furthermore, Figure~\ref{fig:water4} shows that applying Eq.~\ref{DunwegKremer} together with the calculated $\eta$ leads to an estimate of $D(\infty)$, which is consistent with the extrapolated value. This demonstrates that $\eta$ is much less system size dependent than $D(\infty)$, and that the Stokes-Einstein relation, which predicts an inverse relation between these two quantities, is no longer valid on nanometre scale. 

\begin{figure}
  \includegraphics[width=8cm]{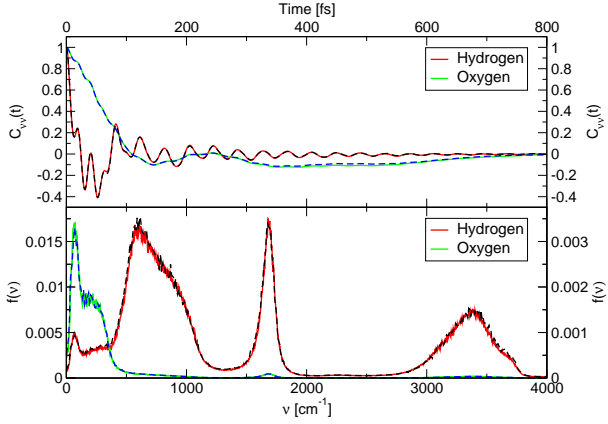}
  \caption{\label{fig:water5} Velocity-velocity autocorrelation function and the corresponding vibrational density of states. The full lines are obtained from second-generation Car-Parrinello simulations, whereas the dashed lines represent BOMD reference calculations.}
\end{figure}

The velocity-velocity autocorrelation function for the hydrogen and oxygen atoms, as well as its temporal Fourier transform that represents the vibrational density of states, is shown in Figure~\ref{fig:water5}. The latter is of particular interest, since, beside being in excellent agreement with our BOMD reference calculations, it provides information about the dynamics of the HB network in liquid water. The small shoulder of the peak in the high-frequency oxygen-hydrogen stretching band indicates only an insignificant presence of dangling HBs, which implies that the time-averaged charge distribution in liquid water is mainly symmetric and the coordination is distorted but tetrahedral.

\textbf{Hydrogen bond kinetics.} The kinetics of HB rearrangements is studied using the Luzar-Chandler model~\cite{a:luzar}, which is able to describe the complex non-exponential relaxation behaviour of HBs with just two rate constants $k$ and $k^{\prime}$ and the following reactive flux correlation function:
\begin{eqnarray}
  k(t) &=& -\frac{dc(t)}{dt} = - \frac{\left< (dh/dt)_{t=0} \left[ 1-h(t) \right] \right>}{\left< h \right>} \nonumber \\
  &=& kc(t) - k^{\prime}n(t), 
  \label{LuzarChandler}
\end{eqnarray}
in which $c(t)={\left< h(0) h(t) \right>}/{\left< h \right>}$ is the HB autocorrelation function, $n(t) = {\left< h(0) \left[ 1 - h(t) \right] H(t) \right>}/{\left< h \right>}$, where H(t) is unity if the molecules of a selected pair are closer than $R=3.5~\AA$ and zero otherwise, while brakets $\left< \cdot \right>$ denote temporal averages. Thus $n(t)$ represent the number of initially bonded pairs, that are broken at time $t$ while remain closer than R. The HB population operator $h(t)$ is defined using geometric descriptors from Ref.~\onlinecite{a:luzar}. The HB lifetime is related to $k$ by $\tau_{\text{HB}} = k^{-1}$, whereas the HB relaxation time is computed as
\begin{eqnarray}
  \tau_{r} = \frac{\int{dt \, tc(t)}}{\int{dt \, c(t)}}.
  \label{HBrelaxation}
\end{eqnarray}

\begin{figure}
  \includegraphics[width=8.0cm]{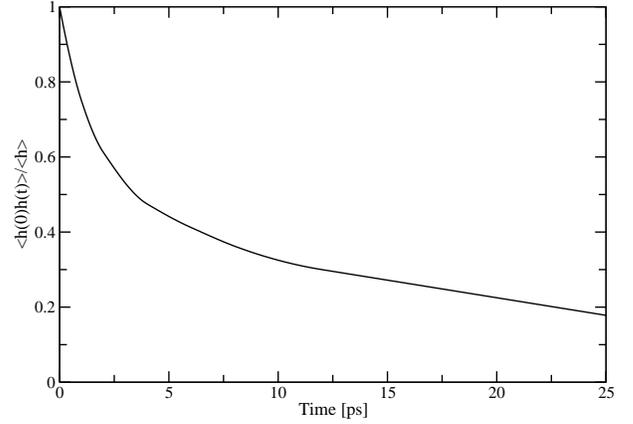}
  \caption{The HB autocorrelation as a function of simulation time.}
 \label{HBautocorrelation}
\end{figure}

The reactive flux correlation function $k(t)$ is non-exponential and monotonically decaying after a period of few librations. A least squares fit of the simulation data obtained with the second-generation Car-Parrinello method to Eq.~(\ref{LuzarChandler}) yields $k$=0.143~$\text{ps}^{-1}$ and $k^{\prime}$=0.389~$\text{ps}^{-1}$, thus $\tau_{\text{HB}}$=6.98~ps and $\tau_{r}$=10.25~ps. Compared to the results obtained with empirical potentials~\cite{a:XuBerne}, our AIMD values for $\tau_{\text{HB}}$ and $\tau_{\text{r}}$ are both about twice as large, whereas the ratio $\tau_{r}/\tau_{\text{HB}}$=1.47 is very close to the value $\sim$~1.5 reported by others~\cite{a:lee2, a:XuBerne}. In addition to these quantitative differences, $c(t)$ obtained from \textit{ab initio} simulations decays significantly slower than predicted by he exponential law (Figure~\ref{HBautocorrelation}). In fact, we suspect that the decay might be biexponential, which would assume a second linear equation for $n(t)$. Most likely this behaviour can be attributed to polarization as well as cooperativity effects, which are suposedly better described by DFT.

\textbf{Summary and further applications.} The computationally efficient second-generation Car-Parrinello method extends AIMD simulations with hundreds of water molecules to the nanosecond timescale. These large-scale simulations allow the re-examination of the contribution of finite-size errors in the calculated structural and dynamical properties of liquid water, as well as the re-assessment of the performance of several commonly used XC functionals. It was found that structural properties are well-converged and reproducible. By contrast, dynamical properties are much less established and the novel simulation approach enables the first calculation of the shear viscosity and the HB relaxation time of liquid water from first principles. 
It is demonstrated that AIMD simulations with classical nuclei and the PBE XC functional predict water to be overstructured and less fluid than the real water, with a shear viscosity within a factor of three from the experimental value. Given the fact that nuclear quantum effects are not included in the simulations it can be concluded that PBE provides a qualitative realistic model for interactions between water molecules in the liquid phase.

In addition to the study reviewed here, the second-generation CPMD has been of great use in a recent investigation of the structure of the water-air interface~\cite{a:kuehnewatersurface} and for the calculation of absolute thermodynamic functions of liquid water from first principles~\cite{a:pascalkuehne}.

\subsection{Electronic signature of the instantaneous asymmetry in the first coordination shell of liquid water}


As just presented, our AIMD simulations are consistent with the conventional wisdom that the local structure of liquid water at ambient conditions is tetrahedral~\cite{a:bernal-fowler,a:stillinger-science-1980,a:saykally-water-review-2010}. Although the thermal motion causes distortions from the perfectly tetrahedral configuration, each molecule in the liquid is bonded, on average, to the four nearest neighbours via two donor and two acceptor bonds~\cite{a:stillinger-science-1980}. This traditional view is based on results from X-ray and neutron diffraction experiments~\cite{a:soper-water-xray,a:thg-xray-water-review,a:thg-xray-water-2003}, vibrational spectroscopy~\cite{a:tokmakoff-geissler-2003,a:geissler-tokmakoff-2005,a:unified-saykally-2005,a:skinner-review-2010}, macroscopic thermodynamics data~\cite{b:water1969,a:stillinger-science-1980,a:saykally-water-review-2010} as well as molecular dynamics simulations~\cite{a:thg-xray-water-review,a:hutter-water-hybrid-dft,a:lee1,a:voth-water-review-2009,a:kuhnewater,a:saykally-water-review-2010}.

Nevertheless, as mentioned above, this traditional picture has recently been questioned based on data from the X-ray absorption, X-ray emission and X-ray Raman scattering experiments~\cite{a:nilsson,a:water-xes-1,a:water-xes-2,a:water-xes-3}. The results of these spectroscopic studies have been interpreted as evidence for strong distortions in the HB network with highly asymmetric distribution of water molecules around a central molecule. It has been suggested that a large fraction of molecules form only two strong HBs: one acceptor and one donor bond~\cite{a:nilsson,a:simulation-xray-nilsson,a:water-xes-1,a:water-xes-2,a:water-xes-3,a:nilsson-pettersson-perspective}. However, the``rings and chains'' structure of liquid water~\cite{a:nilsson}, as well as the inhomogeneous two-state model~\cite{a:water-xes-1,a:water-saxs-1} implied by such an interpretation, have been challenged on many fronts~\cite{a:saykally0,a:unified-saykally-2005,a:saykally1,a:thg,a:thg1,a:water-saxs-thg,a:xas-prendergast-galli,a:water-artacho,a:SchmidtWater,a:saykally-water-review-2010,a:soper-myths,a:car-xas-water} and are a matter of an ongoing debate~\cite{a:saykally0,a:nilsson-comment1,a:saykally-response1,a:unified-saykally-2005,a:soper,a:saykally1,a:water-saxs-1,a:xas-prendergast-galli,a:water-artacho,a:thg,a:thg1,a:simulation-xray-nilsson,a:water-chains-rmc,a:one-more-nilsson,a:water-saxs-thg,a:water-models-2009,a:nilsson-pettersson-perspective,a:saykally-water-review-2010,a:soper-myths,a:car-xas-water}.

In this section, we review a computational study of the energetics and symmetry of local interactions between water molecules in the liquid phase performed with ALMO EDA~\cite{a:water-asym}. As demonstrated for the water dimer, the decomposition of the interaction energy into physically meaningful components provides a deeper insight into the nature and mechanisms of intermolecular bonding than the traditional total-energy electronic structure methods. Applying ALMO EDA to liquid water reveals a significant asymmetry in the strength of the local donor-acceptor contacts. DFT-based AIMD simulations performed using the second-generation Car-Parrinello approach~\cite{a:2ndcpmd, a:kuhnewater} enable us to characterise the geometric origins of the asymmetry, its dynamical behaviour, as well as the mechanism of its relaxation. Furthermore, to address the controversial question of whether it is correct to interpret the X-ray spectra of water in terms of asymmetric structures, we present extensive calculations of its X-ray absorption (XA) spectrum and compare the spectral characteristics of water molecules with different degree of asymmetry. 


\textbf{Computational details.} ALMO EDA was performed for the decorrelated configurations collected from a long 70~ps AIMD trajectory, which was generated by taking advantage of the computational efficiency of the second-generation Car-Parrinello method. AIMD simulations with classical nuclei based on the PBE XC functional~\cite{a:pbe} were performed at constant temperature (300~K) and density (0.9966~g/cm$^3$) for a system containing 128 light water molecules. The results of the previous section show that PBE provides a realistic description of many important structural and dynamical characteristics of liquid water, including the RDFs, self-diffusion and viscosity coefficients, vibrational spectrum, and HB lifetime. However, the PBE water with the classical description of the nuclei is somewhat overstructured suggesting that the degree of the network distortion and fraction of broken bonds may be slightly underestimated.
Therefore, we have explicitly verified that the main conclusions presented here are also valid for snapshots generated with simulations with quantum hydrogen nuclei and based on a water model, which rectifies the main shortcomings of the PBE functional (see Supplementary Information in Ref.~\onlinecite{a:water-asym}).

The MOs in our ALMO EDA calculations were represented by a triple-$\zeta$ Gaussian basis set with two sets of polarization functions (TZV2P) optimized specifically for molecular systems~\cite{a:molopt}. A very high density cutoff of 1000 Ry was used to describe the electron density. The XC energy was approximated with the BLYP XC functional~\cite{a:becke, a:lyp}. The Brillouin zone was sampled at the $\Gamma$-point only and separable norm-conserving pseudopotentials were used to describe the interactions between the valence electrons and the ionic cores~\cite{a:gth, a:hgh, a:krackpp}. Performing ALMO EDA with the HSE06 screened hybrid XC functional~\cite{a:hse06b}, which provides better description of band gaps and electron delocalization effects than BLYP~\cite{a:caravati2, a:caravati2a, a:caravati2b, a:caravati4}, does not change the main conclusions presented below (see Supplementary Information in Ref.~\cite{a:water-asym}).

The XA calculations were performed at the oxygen K-edge using the half-core-hole transition potential formalism~\cite{a:xas-hutter, a:xas-iannuzzi, a:hch-water-review} within all-electron Gaussian augmented plane wave formalism to density functional theory~\cite{a:gapw, a:krackae}. The BLYP XC functional and large basis sets (6-311G** for hydrogen and cc-pVQZ for oxygen atoms) were used to provide an adequate representation of the unoccupied MOs in the vicinity of absorbing atoms~\cite{a:xas-iannuzzi}. A density cutoff of 320 Ry was used to describe the soft part of the electron density. The onset energies of the absorption spectra were aligned with the first $\Delta$SCF excitation energy obtained for each oxygen atom from a separate calculation with the same setup. The spectral intensities were calculated as transition moment integrals in the velocity form. To mimic the experimental broadening, the discrete lines were convoluted with Gaussian functions of 0.2~eV width at half-maximum. The final spectrum was obtained by averaging the convoluted spectra of 9,024 oxygen atoms from 141 AIMD snapshots separated by 500~fs (i.e. 64 oxygen atoms were randomly selected in each snapshot). 

\textbf{Electronic asymmetry and its origins.} Two-body delocalization energy components $\Delta E_{x \rightarrow y}$ defined in Eq.~\ref{eq:edeldec} are the main focus of this study. Each of these terms provide an accurate measure of the perturbation of orbitals localized on a molecule by donor or acceptor interactions with a single neighbour in the bulk system. The donor-acceptor energies are calculated to include cooperativity effects, which are essential for a correct description of the HB network~\cite{a:nedaDFT, a:scortinoWater, a:water-cooperativity0}. Furthermore, the two-body terms are natural local descriptors of intermolecular bonding, which allowed us to analyse the molecular network in liquid water without introducing any arbitrary definitions of a HB~\cite{a:kumar}.

\begin{figure}
\includegraphics*[width=8cm]{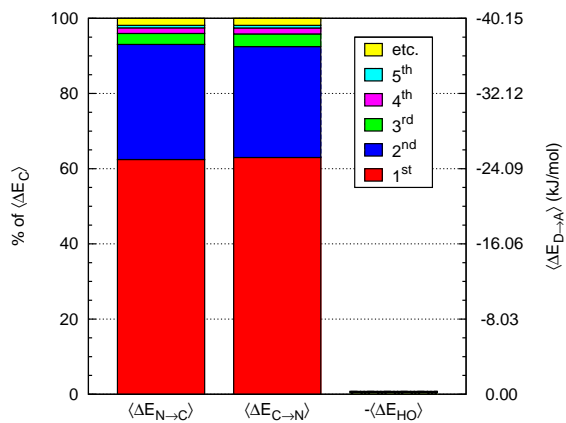}
\caption{\label{fig:del5} Average contributions of the five strongest acceptor $\langle \Delta E_{N \rightarrow C} \rangle$ and donor $\langle \Delta E_{C \rightarrow N} \rangle$ interactions. The rightmost column shows that higher-order delocalization $\langle \Delta E_{HO} \rangle$ does not significantly contribute to the overall binding energy.}
\end{figure}

The electron delocalization energy per molecule $\Delta E_{C}$ can be analysed by neglecting the small higher order relaxation term $\Delta E_{HO}$ (Figure~\ref{fig:del5}) and by considering each water molecule either as a donor or as an acceptor:
\begin{eqnarray}
\label{eq:edelmol}
\Delta E_{C} & \equiv & \sum_{N=1}^{Mol} \Delta E_{C \rightarrow N} = \sum_{N=1}^{Mol} \Delta E_{N \rightarrow C},
\end{eqnarray}
where $C$ stands for the central molecule and $N$ for its neighbours. It is important to emphasise that the terms \emph{donor} and \emph{acceptor} are used here to describe the role of a molecule in the transfer of the electron density. This is opposite to the labelling used for a donor and an acceptor of hydrogen in a HB.

Figure~\ref{fig:del5} shows contributions of the five strongest donor-acceptor interactions to the average delocalization energy of a molecule $\langle \Delta E_{C}\rangle$. Brackets $\langle \ldots \rangle$ denote averaging over all central molecules and AIMD snapshots, which were obtained by performing ALMO EDA for 701 AIMD snapshots separated by 100~fs each (i.e. 89,728 molecular configurations). Figure~\ref{fig:del5} demonstrates that electron delocalization is dominated by two strong interactions, which together are responsible for $\sim$93\% of the delocalization energy of a single molecule. The third and the fourth strongest donor (acceptor) interactions contribute $\sim$5\% and correspond to back-donation of electrons to (from) the remaining two first-shell neighbours (i.e. there is non-negligible delocalization from a typical acceptor to a typical donor). The remaining $\sim$2\% correspond to the delocalization energy to (from) the second and more distant coordination shells.

\begin{figure}
\includegraphics*[width=8cm]{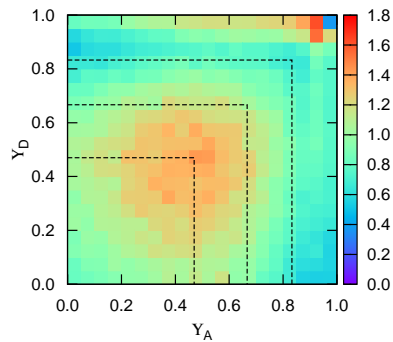}
\caption{\label{fig:asymhist} The normalised probability density function of the asymmetry parameters $\Upsilon_{A}$ and $\Upsilon_{D}$. The probability of finding a molecule in a bin can be found by dividing the corresponding density value by the number of bins (i.e. 400). The dashed black lines at $\Upsilon \approx \frac{1}{2},\frac{2}{3},\frac{5}{6}$ partition all molecules into four groups of equal sizes.}
\end{figure}

Comparison of the strengths of the first and second strongest donor-acceptor interactions ($\sim$25~kJ/mol and $\sim$12~kJ/mol, respectively) with that in the water dimer ($\sim$9~kJ/mol) suggests that each water molecule can be considered to form, on average, two donor and two acceptor bonds. Substantial difference in the strengths of the first and second strongest interactions implies that a large fraction of water molecules experience a significant asymmetry in their local environment. The asymmetry of the two strongest donor contacts of a molecule can be characterised by a dimensionless asymmetry parameter
\begin{eqnarray}
\label{eq:asym}
\Upsilon_{D} = 1 - \frac{\Delta E_{C \rightarrow N^{2nd}}}{\Delta E_{C \rightarrow N^{1st}}}.
\end{eqnarray}
An equivalent parameter $\Upsilon_{A}$ can be used for the two strongest acceptor contacts. The asymmetry parameter is zero if the two contacts are equally strong and equals to one if the second contact does not exist. The probability distribution of molecules according to their $\Upsilon$-parameters is shown in Figure~\ref{fig:asymhist} together with the lines separating the molecules into four groups of equal sizes with different asymmetry. The shape of the distribution demonstrates that most molecules form highly asymmetric donor or acceptor contacts at any instance of time. The line at $\Upsilon \approx 0.5$, for example, indicates that for $\sim$75\% of molecules either $\Upsilon_{A}$ or $\Upsilon_{D}$ is more than 0.5, which means that the strongest donor or acceptor contact is at least two times stronger than the second strongest for these molecules. Furthermore, the peak in the region of high $\Upsilon$ in Figure~\ref{fig:asymhist} indicates the presence of molecules with significantly distorted or even broken HBs.


\begin{figure}
\includegraphics*[width=8.5cm]{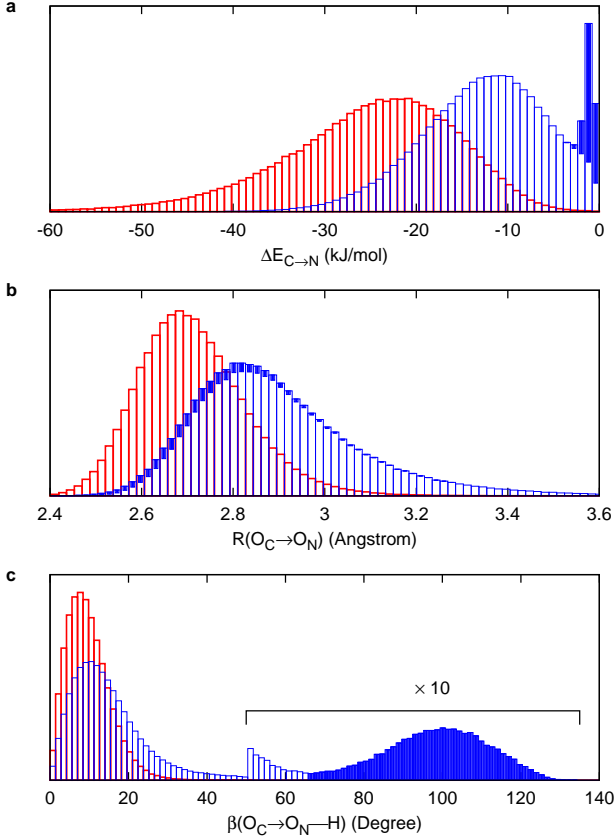}
\caption{\label{fig:hist0} Distribution of the strength (A), intermolecular distance $R$ (B) and HB angle $\beta$ (C) for the first (red) and second (blue) strongest donor interactions $C \rightarrow N$. Filled areas show the contribution of configurations, for which back-donation to a nearby donor is stronger than donation to the second acceptor (cf. text). Distributions for acceptor interactions, $N \rightarrow C$, are almost identical and not shown.}
\end{figure}

Comparison of the configurations of donor-acceptor pairs involved in the first and second strongest interactions reveals the geometric origins of the asymmetry. It has been found that the strength of the interaction is greatly affected by the intermolecular distance $R \equiv d(O_{D} - O_{A})$ and the HB angle $\beta \equiv \angle O_{D}O_{A}H$, while the other geometric parameters have only a minor influence on $\Delta E_{D\rightarrow A}$. The strongly overlapping distributions in Figure~\ref{fig:hist0} suggest that some second strongest interactions have the same energetic and geometric characteristics as the strongest contacts. This implies that the electronic asymmetry observed cannot be attributed to the presence of two distinct types of HBs -- weak and strong. It is, rather, a result of continuous deformations of a typical bond. Another important conclusion that can be made from the distributions in Figure~\ref{fig:hist0} is that relatively small variations of the intermolecular distance ($R\sim 0.2~\AA$) and HB angle ($\beta\sim 5-10^{\circ}$) entails remarkable changes in the strength and electronic structure of HBs. Analysis of the structure of the molecular chains defined by the first strongest bonds (i.e. one donor and one acceptor for each molecule) shows that their directions are random, without any long-range order (i.e. rings, spirals or zigzags) on the length scale of the simulation box ($\sim15~\AA$).
 
The slow decay of the distribution tails (Figure~\ref{fig:hist0}) implies that it is difficult to quantify the concentration of single-donor and single-acceptor molecules in the liquid because defining such configurations using a distance, angle or energy cutoff is an unavoidably arbitrary procedure. A quantitative analysis of the HB network, which was performed in the previous section, shows that, according to the commonly used geometric definitions of HBs~\cite{a:luzar,a:aimd3,a:nilsson}, the structure of water is distorted tetrahedrally with only a small fraction of broken bonds. However, the results presented in Figure~\ref{fig:hist0} indicate that geometric criteria cannot fully characterise the dramatic effect of distortions on the local electronic structure and donor-acceptor interactions of water molecules.  

It is important to note that some second strongest interactions are weakened by distortions to such an extent that back-donation to (from) a nearby donor (acceptor) becomes the second strongest interaction. Such configurations can be clearly distinguished by the large-angle peak in Figure~\ref{fig:hist0}C (the region of the 10-fold magnification). They account for $\sim 6-7$\% of all configurations and are responsible for the low-energy peak in the distribution of $\Delta E_{D \rightarrow A}$ (filled blue areas in Figure~\ref{fig:hist0}) and for the high-$\Upsilon$ peak in Figure~\ref{fig:asymhist}.

\begin{figure}
\includegraphics*[width=8.5cm]{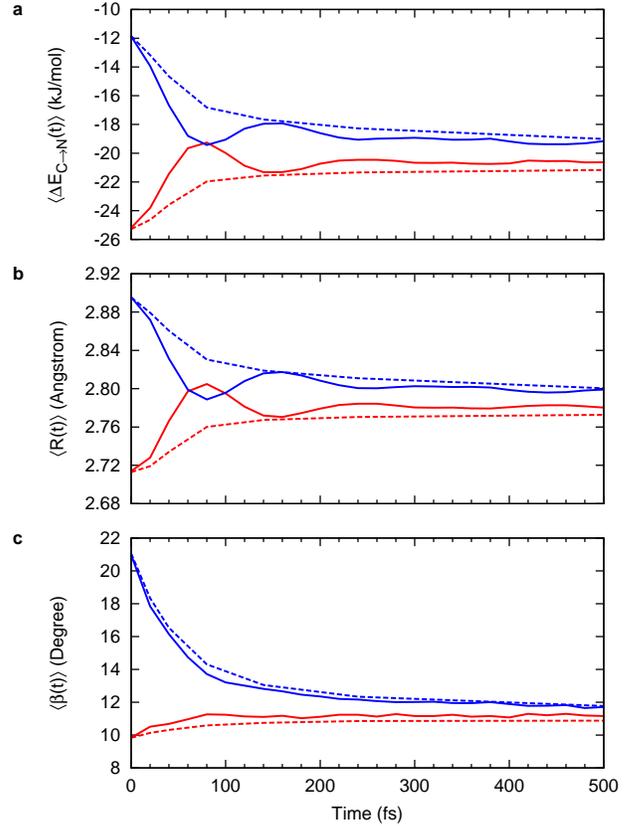}
\caption{\label{fig:trelax} Time dependence of the average strength (A), intermolecular distance $R$ (B) and HB angle $\beta$ (C) for the first (red) and second (blue) strongest donor interactions $C \rightarrow N$. Solid lines show the instantaneous values while the dashed lines correspond to the time-average values. Time-dependent characteristics of acceptor interactions, $N \rightarrow C$, are almost identical and not shown.}
\end{figure}

\textbf{Relaxation of the instantaneous asymmetry.} The overlapping distributions in Figure~\ref{fig:hist0} suggest that, despite the difference in the strength of the donor-acceptor contacts, their nature is similar and the strongest interacting pair can become the second strongest in the process of thermal motion and \emph{vice versa}. To estimate the time scale of this process, it is necessary to examine how the average energy of the first two strongest interactions fluctuates in time. The instantaneous values at time $t$ (solid lines in Figure~\ref{fig:trelax}A) were calculated using the ALMO EDA terms for 3,501 snapshots separated by 20~fs (448,128 local configurations) by averaging over time origins $\tau$ separated by 100~fs and over all \emph{surviving triples}:
\begin{eqnarray}
\label{eq:atime}
& &\langle \Delta E_{C\rightarrow N}(t) \rangle \nonumber \\
& &= \frac{1}{T} \sum_{\tau=1}^{T} \frac{1}{M(\tau,t)} \sum_{C=1}^{M(\tau,t)} \Delta E_{C \rightarrow N}(\tau+t), \end{eqnarray}
where $M(\tau,t)$ is the number of triples that survived from time $\tau$ to $\tau + t$. A triple is considered to survive a specified time interval if the central molecule has the same two strongest-interacting neighbours in all snapshots in this interval.


Figure~\ref{fig:trelax}A shows that the strength of the first two strongest interactions oscillates rapidly and after $\sim$80~fs from an arbitrarily chosen time origin, the first strongest interaction becomes slightly weaker than the second strongest (note that \emph{first} and \emph{second} refer to their order at $t = 0$). The amplitude of the oscillations decreases and the strengths of both interactions approach the average value of $\sim$20~kJ/mol on the timescale of hundreds of femtoseconds. The decay of the oscillations indicates fast decorrelation of the time-separated instantaneous values because of the strong coupling of a selected pair of molecules with its surroundings. In other words, although the energy of a particular HB fluctuates around its average value indefinitely (i.e. with a never-decreasing amplitude), this bond has approximately equal chances of becoming weak or strong after a certain period of time independently of its strength at $t = 0$. This effect is due to the noise introduced by the environment and can be observed in ultrafast infrared spectroscopy experiments~\cite{a:tokmakoff-geissler-2003}.

The time averages shown in Figure~\ref{fig:trelax} are physically meaningful and can be calculated accurately only for the time intervals that are shorter than the average lifetime of a HB $\tau_{\text{HB}} \approx 5-7$~ps, as shown in the previous section~\cite{a:luzar,a:kuhnewater}. The small residual asymmetry that is still present after 500~fs (Figure~\ref{fig:trelax}A) is an indication of the slow non-exponential relaxation behaviour that characterises the kinetics of many processes in liquid water~\cite{a:luzar}.

In addition to the instantaneous values of $\Delta E_{D\rightarrow A}$ \emph{at} time $t$, the dashed lines in Figure~\ref{fig:trelax}A show the corresponding averages \emph{over} time $t$. These values were calculated by averaging over time origins $\tau$, all snapshots lying in the time interval from $\tau$ to $\tau + t$ and over all surviving triples:
\begin{eqnarray}
\label{eq:aovertime}
& &\langle \Delta E_{C \rightarrow N} (t) \rangle^{*} = \nonumber \\
& &= \frac{1}{T} \sum_{\tau=1}^{T} \frac{1}{t+1} \sum_{\kappa=0}^{t} \frac{1}{M(\tau,t)} \sum_{C=1}^{M(\tau,t)} \Delta E_{C \rightarrow N}(\tau+\kappa)
\end{eqnarray}
Time-averages $\langle A(t) \rangle^{*}$ are related to the instantaneous values $\langle A(t) \rangle$ by the following equation:
\begin{eqnarray}
\label{eq:reation}
\langle A(t) \rangle^{*} & \approx & \frac{1}{t} \int_{0}^{t} \langle A(\tau) \rangle d\tau,
\end{eqnarray}
where equality holds if all triples survive over time $t$.

The dashed lines in Figure~\ref{fig:trelax}A shows that any neighbour-induced asymmetry in the electronic structure of a water molecule can be observed only with an experimental probe with a time-resolution of tens of femtoseconds or less. On longer timescales, the asymmetry is destroyed by the thermal motion of molecules and only the average symmetric structures can be observed in experiments with low temporal resolution. 

An examination of the time dependence of all two-body and some three-body geometric parameters that characterise the relative motion of molecules reveals the mechanism of the relaxation. Similar shapes of the curves in Figures~\ref{fig:trelax}A and \ref{fig:trelax}B shows that the relaxation of the asymmetry is primarily caused by low-frequency vibrations of the molecules relative to each other. The minor differences in the behaviour of the curves, in particular at 80~fs, indicate that the relaxation of the asymmetry is also influenced by some other degrees of freedom. The temporal changes in the HB angles towards the average value (Figure~\ref{fig:trelax}C) show that librations of molecules play this minor role in the relaxation process.

The kinetics and mechanism of the asymmetry relaxation presented here are supported by data from ultrafast infrared spectroscopy, which can directly observe intermolecular oscillations with a period of 170~fs~\cite{a:tokmakoff-geissler-2003}. They are also in agreement with the theoretical work of Fernandez-Serra \emph{et al.}, who have used the Mulliken bond order parameter to characterise the connectivity and dynamical processes in the HB network of liquid water~\cite{a:water-artacho}. 

\begin{figure}
\includegraphics*[width=8.5cm]{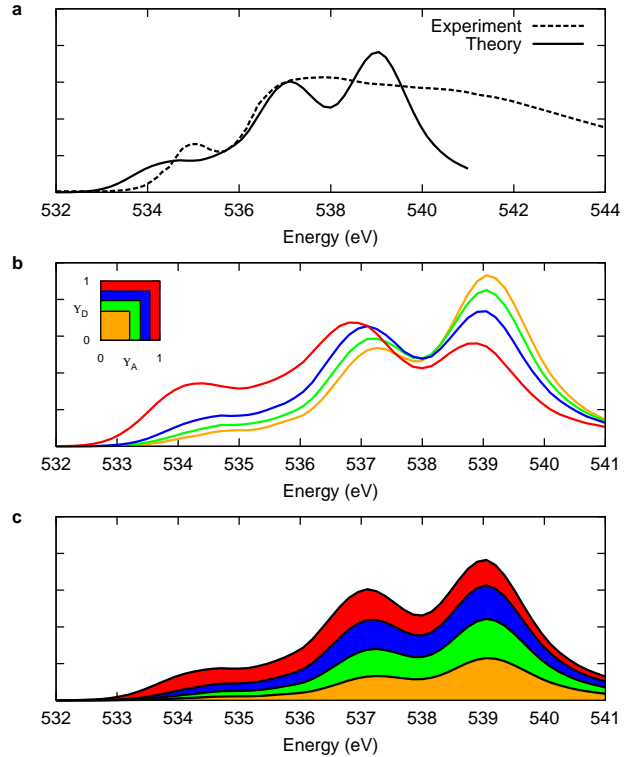}
\caption{\label{fig:xas} A. Calculated and experimental~\cite{a:nilsson} XA spectra of liquid water. B. Calculated XA spectra of the four groups of molecules separated according to the asymmetry of their donor ($\Upsilon_{D}$) and acceptor ($\Upsilon_{A}$) environments as shown in the inset. C. Contributions of the four groups into the total XA spectrum. The colour coding is shown in the inset above.}
\end{figure}

\textbf{X-ray absorption signatures of asymmetric structures.} The time behaviour described above implies that the instantaneous asymmetry can, in principle, be detected by X-ray spectroscopy, which has temporal resolution of several femtoseconds and is highly sensitive to perturbations in the electronic structure of molecules~\cite{a:nilsson,a:nilsson-pettersson-perspective}. To identify possible relationships between the spectroscopic features and asymmetry, the XA spectrum of liquid water is calculated at the oxygen K-edge. 
Although the employed computational approach overestimates the intensities in the post-edge part of the spectrum and underestimates the pre-edge peak and overall spectral width~\cite{a:xas-iannuzzi}, it provides an accurate description of the core-level excitation processes and semi-quantitatively reproduces the main features of the experimentally measured spectra (Figure~\ref{fig:xas}A). 

The localized nature of the $1s$ core orbitals allows the disentanglement of spectral contributions from molecules with different asymmetry. To this end, all molecules are separated into four groups according to the asymmetry of their donor and acceptor environments, as shown in Figure~\ref{fig:asymhist}. Choosing boundaries at $\Upsilon \approx \frac{1}{2},\frac{2}{3},\frac{5}{6}$ distributes all molecules into four groups of approximately equal sizes (i.e. $25\pm 2\%$). Figure~\ref{fig:xas}B shows four XA spectra obtained by averaging the individual contributions of molecules in each group. It reveals that molecules in the symmetric environment exhibit strong post-edge peaks, while molecules with a high asymmetry in their environment are characterised by the amplified absorption in the pre-edge region. Furthermore, the relationship between the asymmetry and absorption intensity is non-uniform: the pre-edge peak is dramatically increased in the spectrum for the 25\% of molecules in the most asymmetric group, for which the first strongest interaction is more than six times stronger than the second. As a consequence, the pre-edge feature of the calculated XA is dominated by the contribution of molecules in the highly asymmetric environments (Figure~\ref{fig:xas}C).

The pronounced pre-edge peak in the experimentally measured XA spectrum of liquid water has been interpreted as evidence for its ``rings and chains'' structure, where $\sim$80\% of molecules have two broken HBs~\cite{a:nilsson,a:simulation-xray-nilsson}. The results presented here suggest that this feature of the XA spectrum can be explained by the presence of a smaller fraction of water molecules with high instantaneous asymmetry. Although the employed XA modelling methodology does not allow the precise estimation of the size of this fraction, this conclusion is consistent with that of recent theoretical studies at an even higher level of theory, which have demonstrated that the main features of the experimental XA spectra can be reproduced in simulations based on conventional nearly-tetrahedral models~\cite{a:car-xas-water,a:kong-xas}. Thus, the presented application of ALMO EDA to liquid water complements the previous results by revealing an interesting and important connection between relatively small geometric perturbations in the hydrogen-bond network, the large asymmetry in the electronic ground state and the XA spectral signatures of the core-excitation processes.

\textbf{Summary.} The main findings of our investigation of local donor-acceptor interactions of water molecules with their neighbours in the liquid phase are as follows:
\begin{itemize}
\item The strength of donor-acceptor interactions suggest that each molecule in liquid water at ambient conditions forms, on average, two donor and two acceptor bonds.
\item Even small thermal distortions in the tetrahedral HB network induce a significant asymmetry in the strength of the contacts causing one of the two donor (acceptor) interactions to become, at any instance of time, substantially stronger than the other. Thus, the instantaneous structure of water is strongly asymmetric only according to the electronic criteria, not the geometric one.
\item Intermolecular vibrations and librations of OH groups of HBs result in the relaxation of the instantaneous asymmetry on the timescale of hundreds of femtoseconds.
\item The pronounced pre-edge peak observed in the XA spectra of liquid water can be attributed to molecules in asymmetric environments created by instantaneous distortions in the fluctuating but on average symmetric HB network.
\end{itemize}

\section{Conclusion}

We conclude by noting that the development of second-generation CPMD~\cite{a:2ndcpmd} as well as ALMO EDA~\cite{a:theeda,a:water-asym} are both for themselves important methodological achievements that had been instrumental to elucidate the water dimer and liquid water, respectively. 
Nevertheless, neither method alone but only the combination allowed for the additional insights to eventually reconcile the two existing and seemingly opposite models of liquid water -- the traditional symmetric and the recently proposed asymmetric -- and represent an important step towards a better understanding of the electronic structure of the HB network of one of the most important liquids on Earth.


\begin{acknowledgments}
R.Z.K. is grateful to the Swiss National Science Foundation for financial support and to the Swiss National Supercomputing Centre (CSCS) for computer time. The authors would also like to thank the Graduate School of Excellence MAINZ as well as the IDEE project of the Carl Zeiss Foundation for financial support.
\end{acknowledgments}

%

\end{document}